\documentclass[aps,pra,showpacs,twocolumn,superscriptaddress]{revtex4}
\usepackage{amsmath}
\usepackage{amssymb}
\usepackage{epsfig}
\usepackage[utf8x]{inputenc}
\usepackage{hyperref}
\usepackage{color}
\definecolor{citekey}{rgb}{0,1,1}
\definecolor{refkey}{rgb}{1,0,0}
\definecolor{labelkey}{rgb}{0,0,1}

\graphicspath{{.}{./EPS/}}

\newcommand* {\ee}{\ensuremath{\mathrm{e}}}
\newcommand{\ket}[1]{|{#1}\rangle}
\newcommand{\bra}[1]{\langle{#1}|}

\newcommand{\braket}[2]{\langle{#1}|{#2}\rangle}

\begin{document}
 
\title{
Quantum teleportation and entanglement swapping of matter 
qubits with coherent multiphoton states 
}

\author{J. M. Torres}
\email{Mauricio.Torres@physik.tu-darmstadt.de}
\affiliation{Institut f\"{u}r Angewandte Physik, Technische Universit\"{a}t Darmstadt, D-64289, Germany}
\affiliation{Departamento de Investigaci\'on en F\'isica, Universidad de Sonora, Hermosillo, M\'exico}
\author{J. Z. Bern\'ad}
\email{Zsolt.Bernad@physik.tu-darmstadt.de}
\affiliation{Institut f\"{u}r Angewandte Physik, Technische Universit\"{a}t Darmstadt, D-64289, Germany}
\author{G. Alber}
\email{gernot.alber@physik.tu-darmstadt.de}
\affiliation{Institut f\"{u}r Angewandte Physik, Technische Universit\"{a}t Darmstadt, D-64289, Germany}

\date{\today}

\begin{abstract}
Protocols for probabilistic entanglement-assisted quantum teleportation and for entanglement
swapping of material qubits are presented. They are based on a protocol for postselective Bell-
state projection which is capable of projecting two material qubits onto a Bell state with the help
of ancillary coherent multiphoton states and postselection by balanced homodyne photodetection.
Provided this photonic postselection is successful we explore the theoretical possibilities of realizing unit fidelity quantum teleportation
and entanglement swapping with $25\%$ success probability. This photon-assisted Bell projection is generated by coupling
almost resonantly the two material qubits to single modes of the radiation field in two separate cavities
in a Ramsey-type interaction sequence and by measuring the emerged field states in a balanced homodyne detection scenario.
As these quantum protocols require basic tools of quantum state engineering
of coherent multiphoton states and balanced homodyne photodetection they may offer interesting
perspectives in particular for current quantum optical applications in quantum information processing.
\end{abstract}

\maketitle

\section{Introduction}
The development of physical procedures for 
the establishment of entanglement between distant material quantum systems, such as qubits,
capable of storing quantum information reliably
is an important prerequisite for quantum communication \cite{Nielsen}.
Such material quantum systems may form the
nodes of a quantum network \cite{Kimble}, for example,
which are possibly also connected
by photonic channels enabling the direct transfer of quantum information
or the establishment of entanglement.
However, as typically direct transfer of quantum information over photonic channels
is affected by loss processes and by decoherence it may be advantageous
to exploit already existing entanglement between nodes within such a network
for purposes of reliable exchange 
of quantum information. Furthermore, controlled redistribution of
entanglement within such a quantum network may be used to establish new
routes for exchange of quantum information.
Reliable
transfer of quantum information may be achieved with the help of
entanglement-enabled quantum teleportation \cite{Bennett93}
and redistribution of entanglement with the help of entanglement swapping.
In order to be able to realize these two
important elementary quantum information processing protocols 
in material qubit systems it is necessary to implement projective Bell measurements which
can be performed reliably locally at each node of such a quantum network.
Complete Bell measurements capable of distinguishing all four
Bell states are still difficult to realize experimentally. In view of these considerable
experimental difficulties it is of current interest to
develop implementations of perfect postselective
Bell projections. In such a projective Bell measurement two
material qubits are projected onto a particular Bell
state probabilistically in such a way that,
provided this projective measurement is successful,
this two-qubit Bell state is postselected with unit fidelity.

Recently, several proposals have been made for implementing a quantum repeater
\cite{repeaterreview}
which redistributes entanglement from intermediary entangled material qubit pairs
to distant qubits with the help of entanglement swapping
\cite{Zuk, Riedmatten05, Halder07, Xue12}.
Thereby imperfections affecting the entanglement swapping can be compensated
afterwards by entanglement purification \cite{Dur99,Zhao03,Reich06}.

First physical 
implementations of entanglement-assisted
quantum teleportation were realized with
photonic qubits \cite{Boschi98, Bouwmeester97, Furusawa98}. Subsequent
experiments achieved teleportation over 
distances of $100\, {\rm km}$ \cite{Yin12, Ma12}. 
First experiments on teleportation with material qubits
were limited to distances of the order of a
few ${\rm \mu m}$ \cite{Riebe04, Barrett04}. However, 
most recent realizations
report successful teleportation with material qubits
over distances of $21\, {\rm m}$ \cite{Nolleke13} with the help of
ancillary photon exchange.

Despite these recent experimental advances
these realizations of quantum information
transfer are limited to distances of the order of $100\, {\rm km}$ 
mainly due to the use of single or few photon states acting as ancillary quantum systems.
In order to overcome this hurdle coherent photon states
offer interesting perspectives. Techniques for their
generation, manipulation and detection 
are well developed and
these multiphoton states of the radiation field
can be transmitted in a controlled way through already existing
optical communication networks.
The hybrid quantum repeater model of van Loock et al. 
\cite{vanLoock} is an early example which
aims at exploiting these advantages
of coherent multiphoton states for purposes of quantum 
information processing.

Motivated by these advantages
and by the fundamental role played by Bell-state projections in 
basic quantum communication protocols 
in this paper we propose a protocol for implementing a
probabilistic Bell-state projection of material qubits with the help
of coherent multiphoton states and 
of photonic postselection by
balanced homodyne photodetection. 
This postselective measurement
protocol results in
a Bell state with almost unit fidelity and success probability 
depending on the overlap of the initial material state with this specific Bell state.
In our scenario 
single modes of the radiation field initially
prepared in coherent states are 
used as ancillary quantum systems in a Ramsey-type interaction sequence. These
photonic states have a specific phase difference and 
interact almost resonantly
with the two qubits 
for appropriately chosen interaction times. 
Built on this procedure we propose 
probabilistic protocols for entanglement-assisted quantum teleportation
and for entanglement swapping. 

The probabilistic photon-assisted
Bell projection discussed in this paper
is based on two crucial dynamical properties.
Firstly, it takes advantage of a characteristic property
of the two-qubit
Tavis-Cummings model \cite{Tavis} describing
the almost resonant interaction between two qubits
and a single mode of the radiation field, namely
the existence of an invariant two-qubit Bell 
state which is not coupled to the photons.
However, this characteristic
property with the aid of a photonic postselection
can generate an almost perfect
Bell state only for  specific initial conditions.
In the case of arbitrary initial conditions 
the
two-qubit quantum states resulting from a photonic postselection
are noisy Bell states. 
It is shown that the elimination of these noisy contributions
can be achieved by the second essential
dynamical property of our protocol, namely the involvement of a
Ramsey-type interaction scenario.

This paper is organized as follows.
In Sec.\ref{TavisSolution}
the quantum electrodynamical interaction between two 
material qubits and a single mode of the radiation field is discussed
within the framework of the Tavis-Cummings \cite{Tavis} model.
Approximate analytical solutions are presented for the time
evolution of the entangled matter-photon quantum state which
are valid for initially prepared coherent field states and
for almost resonant interaction between the two qubits and the photons. 
Furthermore, a detailed discussion of the two-qubit
quantum state is presented which results from photonic 
postselection by balanced
homodyne detection.
In Sec. \ref{Filter} these results are generalized
to a Ramsey-type interaction scenario involving two 
subsequent matter-field interactions in two cavities and two
photonic postselection processes by balanced homodyne detection.
It is shown that this procedure can prepare a Bell state with
unit fidelity for any given initial condition of the two material qubits
and with success probability given by the initial probability weight 
of the generated Bell state.
In Sec. \ref{stability} we discuss effects that arise
from unequal coupling strengths of the qubits  to the radiation field
and from different transition frequencies of the qubits.
Finally, in Sec.
\ref{QSchemes} 
implementations of entanglement-assisted
quantum teleportation and entanglement swapping 
are discussed which are based on the postselective Bell-state projection 
of Sec. \ref{Filter}.
A detailed derivation of the solution of the two-qubit
Tavis-Cummings model is given in Appendix \ref{Appsol}. 
For the sake of completeness in Appendix \ref{homodynedetection}
basic facts concerning the theoretical description of
balanced homodyne photodetection are summarized. In Appendix \ref{DiffAt}
we include analytical calculations that support Sec. \ref{stability}.

\section{The two-qubit Tavis-Cummings model}
\label{TavisSolution}
   
In this section we discuss basic dynamical features of the 
two-qubit Tavis-Cummings model \cite{Tavis}. This model describes the interaction between 
two two-level systems (material qubits) and
a single-mode of the radiation field inside a cavity. 
As this model involves an interaction-insensitive two-qubit Bell state it is
possible to prepare this maximally entangled two-qubit state by projection onto an appropriate
photonic quantum state. For initially prepared coherent states of the radiation field this projection
can be achieved by postselection with the help of 
balanced homodyne photodetection.

\subsection{The qubit-field dynamics}
We consider two
two-level systems (material qubits), say  $A$ and $B$,
with ground states $\ket{0}_i$ and excited states 
$\ket{1}_i$ ($i \in\{A,B\}$) separated by an energy difference
$\hbar \omega_{\rm a}$ from their
ground states. 
Both two-level systems are assumed to have equal transition dipole moments between the almost
resonantly coupled energy eigenstates $\ket{0}_i$ and $\ket{1}_i$ of different parity. 
In the dipole and rotating-wave approximation 
the two-qubit Tavis-Cummings Hamiltonian describing almost resonant interaction
of these two qubits with a single mode of the radiation field is given by
\begin{align}
  \hspace{-.23cm}\hat{H}&= 
\hbar \omega \hat{a}^\dagger \hat{a} 
+\hspace{-.21cm}\sum_{i=A,B}\hspace{-.13cm}\hbar\left( 
\frac{\omega_{\rm a}}{2}\hat\sigma_i^z
+ g  \ee^{i\theta}\hat{\sigma}^+_i\hat{a}+ g 
 \ee^{-i\theta} \hat{\sigma}^-_i\hat{a}^\dagger\right)
\label{Hamilton}
\end{align}
with 
 $\hat\sigma_i^z=\ket{1}\bra{1}_i-\ket{0}\bra{0}_i$
 ($i \in\{A,B\}$).
The ladder operators of the qubits are denoted by
$\hat{\sigma}^+_i=\ket{1}\bra{0}_i$ and  $\hat{\sigma}^-_i=\ket{0}\bra{1}_i$
and
the radiative coupling of the qubits to the single mode of the radiation field is
characterized by the
vacuum Rabi frequency  $2g$ and the phase $\theta$. 
The annihilation and creation operators
of the single-mode radiation  field with frequency $\omega$ 
are denoted by
$\hat{a}$  and $\hat{a}^\dagger$. The detuning between the radiation field
and the transition frequency of the two-level systems is given by
$\delta = \omega_a - \omega$.

In our subsequent discussion we are particularly interested in
solutions of the time dependent Schr\"odinger equation governed by
the Hamiltonian of Eq. \eqref{Hamilton}. We assume that initially
the matter-field system
is prepared in a pure separable quantum state 
\begin{align}
  \ket{\Psi_0}=&\left(
  c_-\ket{\Psi^-}+
  c_1\ket{1,1}+c_+\ket{\Psi^+}+
  c_0\ket{0,0}
  \right)\ket{\alpha},
  \label{initial}
\end{align}
where the pure two-qubit state is expanded in the
orthonormal  Bell states
\begin{align}
  \ket{\Psi^\pm}&=\tfrac{1}{\sqrt2}(\ket{0,1}\pm\ket{1,0})
  \label{bellpsi}
\end{align} 
and the separable states $|1,1\rangle$ and $|0,0\rangle$ with 
$\ket{i}_A\ket j_B=\ket{i,j}$ ($i,j \in \{0,1\}$). The single mode of the radiation field is in a coherent state 
\begin{align}
  \ket{\alpha}=\sum_{n=0}^\infty 
  \ee^{-\frac{|\alpha|^2}{2}}
 \frac{\alpha^n}{\sqrt{n!}}
 \ket n,
  \quad\alpha=\sqrt{\overline n}\,\ee^{i\phi}
  \label{coherentstate}
\end{align}
with phase $\phi$, mean photon number $\overline n$ and $\ket{n}$($n\in {\mathbb N}_0$) denoting the normalized photon-number states. Normalization of $\ket{\Psi_0}$ requires the condition
$|c_-|^2+|c_+|^2+|c_0|^2+|c_1|^2=1$.

In the following we shall take advantage
of a special feature of the two-qubit Tavis Cummings model, namely that
quantum states of the form 
$\ket{\Psi^-}\ket{n}$($n\in {\mathbb N}_0$) with the photon-number state $\ket n$ 
are stationary eigenstates of the Hamiltonian of Eq. \eqref{Hamilton} with eigenvalue
$\hbar\omega n$.  

The time evolution of an initial state of the form of  Eq. \eqref{initial}
can be obtained from the solution of
the eigenvalue problem of the two-qubit Tavis-Cummings Hamiltonian \eqref{Hamilton}.
This solution is presented in detail in Appendix \ref{Appsol}. 
Here, we merely present the final result of the time dependent tripartite quantum state, i.e.
\begin{align}
  \ket{\Psi(t)}&=
  c_-\ket{\Psi^-}\ket{\alpha \ee^{-i\omega t}}+
  \nonumber \\&
  +\ket{1,1}\ket{\chi_{1}(t)}
  +\ket{\Psi^+}\ket{\chi_{0}(t)} 
  +\ket{0,0}\ket{\chi_{-1}(t)}.
  \label{psi}
\end{align}
The  matter-field state $\ket{\Psi(t)}$ displays the interaction between the material
systems $A$, $B$ and the single-mode of the radiation field.
According to Eq. \eqref{psi}  the coherent state 
$\ket{\alpha \ee^{-i\omega t}}$ is strictly 
correlated with the maximally entangled material Bell state
$\ket{\Psi^-}$.  Therefore, if we were able to discriminate the field state
$\ket{\alpha \ee^{-i\omega t}}$ from the other three  field states
$\ket{\chi_j(t)}$ $\left(j\in\{-1,0,1\}\right)$ we could prepare the
maximally entangled material Bell state $\ket{\Psi^-}$ in a probabilistic way.
However, this discrimination of the field states  is not a straightforward task as
they are not orthogonal, in general, so that they cannot be distinguished perfectly.
For our subsequent development of a probabilistic scheme for entanglement swapping and
quantum teleportation based on coherent field states and photonic postselection by
homodyning it will be of crucial importance to be able to distinguish these field states
almost perfectly.

\begin{figure}
  \includegraphics[width=.49\textwidth]{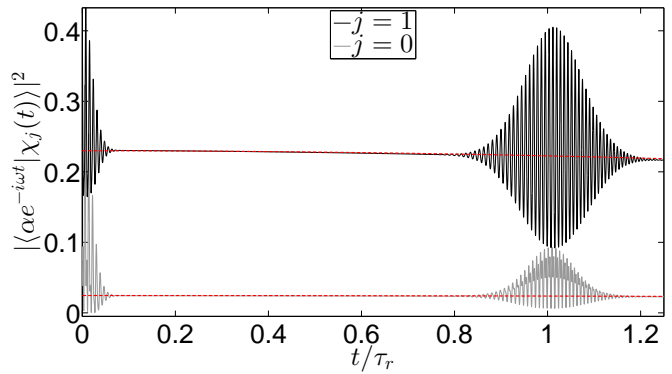}
  \caption{\label{overlap1}
  Overlap between the exact photonic states $\ket{\chi_j(t)}$ (see Eq. \eqref{psi})
  and the coherent state $\ket{\alpha \ee^{-i\omega t}}$ showing the collapse 
  and revival phenomena. 
  Typically, in the collapse
  region the overlap is nonzero and proportional to the parameter
  $|\eta(\vec c,\phi)|^2$ of 
  Eq. \eqref{etas}.
  The upper and lower 
  lines (red) show the approximation given in Eq.\eqref{overlapplateau}.
  The overlap for $j=-1$ is not shown and 
  behaves qualitatively as for $j=1$.
  The parameters  are $\alpha=7.6\ee^{i2.65}$, 
$c_-=0.5446 \ee^{i },\,c_1=0.6389 \ee^{-i1.8},\,c_+=0.1950 \ee^{-i0.3},\,c_0=0.5071 \ee^{i1.3}
$,  $\delta/g=3.5$, and $\theta=0$. }
\end{figure}

Some basic properties of the pure field states  
which determine the tripartite quantum state
$\ket{\Psi(t)}$ can be studied by considering the time dependence of the overlaps
between the coherent state $\ket{\alpha \ee^{-i\omega t}}$
and the other three field states 
$\ket{\chi_j(t)}$ $(j\in\{-1,0,1\})$
as depicted in 
Fig. \ref{overlap1}. These overlaps
resemble  collapse and revival phenomena which also appear in a similar form
in the Jaynes-Cummings model \cite{Schleich}. 
After a collapse time $\tau_c$ initial rapid oscillations of the overlaps decay to a
'plateau' characterized by an almost constant value. After a revival time $\tau_r$
the rapid oscillations reappear. Thus, a perfect discrimination of the material Bell state
$\ket{\Psi^-}$ from the other material quantum states $\ket{0,0}$ and $\ket{1,1}$ would be possible
in the plateau region if these overlaps vanished. However, such vanishing overlaps in the plateau region
can only be achieved for very special initial conditions of the two qubits
as will be demonstrated in the following.

In order to gain insight into the intricate dynamical evolution of $\ket{\Psi(t)}$ let us concentrate on the case of large mean 
photon numbers.  For initial field states $\ket{\alpha}$ with  $\overline{n}\gg 1$
it is possible to simplify the time dependent solution $\ket{\Psi(t)}$
of Eq.\eqref{psi} significantly
by expanding the eigenvalues of the Tavis-Cummings Hamiltonian around $\overline{n}$ up to
first order in $n$, i.e.
\begin{align}
  E^{(n)}_j&\approx
  \hbar\left[\Delta_j+\left(\omega+\varpi_j\right)\left(
  n-1
  \right)\right]
  \label{eigenvalues}
\end{align}
with 
\begin{align}
  \Delta_j&=j\frac{2g^2\overline n+\delta^2}{\Omega_{\overline n}}
  +
  \delta 
  g^2\frac{\omega_{\overline n}^2+\Omega_{\overline n}^2-2g^2}{\Omega_{\overline n}^4}
  (-1)^j2^
  {\delta_{j,0}},
  \nonumber\\
  \varpi_j&=
  j\frac{2g^2}{\Omega_{\overline n}}
  -\frac{4\delta}{(\Omega_{\overline n} /g)^4}
  (-1)^j2^
  {\delta_{j,0}},
\nonumber\\
  \omega_n&=g\sqrt{4n-2},\quad \Omega_n=\sqrt{g^2(4n-2)+\delta^2}
  \label{frequenciesTaylor}
\end{align}
and with  the Kronecker delta $\delta_{i,j}$.
The index $j=-1,0,1$ distinguishes the three eigenvalues of each coupled block with
photon number $n$.
Depending on whether $|j|=1$ or $j=0$
 the frequency  $\Delta_j$ introduces 
two largely different time scales because in the limit $\overline{n}\gg 1$ we obtain the result
$\Delta_{\pm 1}/\Delta_0\sim\varpi_{\pm 1}/\varpi_{0}\sim3\overline n$.
According to the first order expansion the validity of Eq. \eqref{eigenvalues} is restricted to times $\tau$ with
\begin{equation}
  \frac{1}{2\hbar}\left| \frac{d^2E_j^{(n)}}{dn^2}\right|_{n=\overline{n}}
\tau  \overline{n} \ll 2 \pi. 
\label{timescale} 
\end{equation}
In this approximation
the field states can be written as a superposition of coherent states , i.e.
\begin{equation}
  \ket{\chi_{j}(t)}\hspace{-.1cm}=
 \hspace{-.25cm}\sum_{k=-1}^1
  \hspace{-.15cm}\eta_{j,k}
\ee^{i\left[j(\phi+\theta-(\omega+\varpi_k) t)-\Delta_kt\right]}
  \ket{\alpha \ee^{-i(\omega+\varpi_k)t}}
  \label{chistatesdelta}
\end{equation}
with the parameters 
\begin{align}
  &\eta_{j,0}
  =
  \left(\frac{\delta}{\omega_{\overline n}}\right)^
  {\delta_{j,0}}\frac{(-1)^{\delta_{j,1}}}{\sqrt{2^{|j|}}}\,
  \eta(\vec c,\phi),\quad \vec c=(c_+,c_0,c_1),
  \label{etas}
  \nonumber\\
  &\eta (\vec c,\phi) =
  \frac{\omega_{\overline{n}}^2}{\Omega_{\overline{n}}^2}
  \left(
  \frac{\delta}{\omega_{\overline{n}}} c_+
  +
  \frac{
  c_0\ee^{i(\phi+\theta)}-c_1 \ee^{-i(\phi+\theta)}}
  {\sqrt2}
  \right),
  \nonumber\\
  &\eta_{j,\pm1}
  =
  \left(
  \frac{\delta\pm\Omega_{\overline{n}}}{\omega_{\overline{n}}}
  \right)^{j}
  \frac{\omega_{\overline{n}}^2}{\sqrt{2^{|j|}}2\Omega_{\overline{n}}^2}
   \times
  \nonumber\\
  &
  \quad\quad
  \left(
  c_++
  \frac{
  \omega_{\overline{n}}\,c_0 \ee^{i(\phi+\theta)}}
  {\sqrt{2}(\delta\pm\Omega_{\overline{n}})}+
  \frac{
  (\delta\pm\Omega_{\overline{n}})\,c_1 \ee^{-i(\phi+\theta)}}
  {\sqrt{2}\omega_{\overline{n}}}
  \right).
\end{align}
These approximate solutions of the field states yield further insight into the
collapse and revival phenomena apparent in Fig. \ref{overlap1} as these overlaps
are determined by 
\begin{align}
  \braket{\alpha \ee^{-i\omega t}}{\alpha \ee^{-i(\omega+\varpi_j)t}}&
  =\ee^{-\bar n(1-\ee^{-i\varpi_j t})}.
  \nonumber\\&
\approx \ee^{-\bar n(i\varpi_j t+{\varpi_j}^2 t^2/2)}.
\label{overlap}
\end{align}
An additional approximation of the last line is valid only for short times $t$ with $t\ll 2\pi/\varpi_j$. 
In order to meet the requirement of condition \eqref{timescale}
we have to restrict our description to the shortest time scale or highest frequencies
$\varpi_{\pm1}$.
These two frequencies are of the same order 
and characterize time scales of the collapse and the revival
phenomena.
The revival time $\tau_r$ is characterized by a vanishing exponent in the first line of Eq.
\eqref{overlap}. 
The exponential decay in the second line defines the collapse time $\tau_c$.
Accordingly, these two characteristic times are given by
\begin{align}
  \tau_r=\frac{\pi}{g}\sqrt{4\overline n-2+\frac{\delta^2}{g^2}}, \quad 
  \tau_c=\frac{\tau_r}{\pi\sqrt{2\overline n}}.
  \label{revcoltime}
\end{align}
Therefore, for 
interaction times $\tau$ in the plateau region of Fig. \ref{overlap1}, i.e.
$  \tau_c<\tau\ll\tau_r $
the relevant overlaps between the  field states
can be approximated by
\begin{align}
  |\braket{\alpha \ee^{-i\omega \tau}}{\chi_j(\tau)}|^2
  =
  \tfrac{\ee^{-\overline n \varpi_0^2\tau^2 }}{2^{|j|}}
  \left(\tfrac{\delta}{\omega_{\overline n}}\right)^{2\delta_{j,0}}
   |\eta(\vec c,\phi)|^2. 
  \label{overlapplateau}
\end{align}
From Eq. \eqref{overlapplateau} it is apparent that in the plateau region the three relevant overlaps are proportional
to the parameter $\eta(\vec c,\phi)$ 
of Eq. \eqref{etas}. The overlap between $\ket{\alpha \ee^{-i\omega t}}$ and
$\ket{\chi_0(t)}$ is the only one which is proportional to the detuning $\delta$.
Therefore, for  interaction times $\tau_c \ll \tau \ll \tau_r$
the state $\ket{\chi_0(t)}$ is always orthogonal
to the free coherent 
state $\ket{\alpha \ee^{-i\omega t}}$ provided the interaction between the two-level systems and the single mode of the radiation field is
resonant, i.e. $\delta=0$. 

Let us now consider a projective field measurement of the coherent state $\ket{e^{-i\omega \tau}\alpha }$.
The time evolution of the tripartite system is 
given by Eqs. \eqref{psi} and  \eqref{chistatesdelta}.
For interaction times in the plateau region of Fig. \ref{overlap1} , i.e.
$\tau_c \ll \tau \ll \tau_r$,  we obtain 
as a result of such a projective field measurement
the unnormalized two-qubit quantum state 
\begin{align}
  \bra{\alpha \ee^{-i\omega \tau}}
  \ee^{-i{\hat H}\tau/{\hbar}}
  \ket{\Psi_0}=
  c_-\ket{\Psi^-}+
  \eta (\vec c,\phi)s\ket{\psi_\phi}
  \label{projection}
\end{align}
with
\begin{align} 
s=\ee^{-i(\Delta_0+\varpi_0\overline n)\tau-\overline n\varpi_0^2\tau^2/2}.
\label{projection2}
\end{align}
This material quantum state
is a superposition of the antisymmetric Bell state $\ket{\Psi^-}$
and the unnormalized state
\begin{align}
  \ket{\psi_\phi}&=
  \frac{\delta}{\omega_{\overline n}}\ket{\Psi^+}
  -\frac{\ee^{i(\Theta+\phi)}}{\sqrt2}\ket{1,1}
  +\frac{\ee^{-i(\Theta+\phi)}}{\sqrt2}\ket{0,0},
\label{undestate}
\end{align}
where we have introduced the phase  
$\Theta=\theta-(\omega+\varpi_0)\tau$.
The parameter $\eta(\vec c,\phi)$ is given by Eq. 
\eqref{etas}.
The normalization of the state after the projection as given by Eq. \eqref{projection}
yields the success probability $P$ of the projective field measurement, i.e.
\begin{align}
  P=|c_-|^2+|\eta(\vec c,\phi)|^2\left(1+\frac{\delta^2}{\omega_{\overline n}^2}\right) 
\ee^{-\overline n\varpi_0^2\tau^2}.
  \label{Prob}
\end{align}

Thus, perfect projection onto the antisymmetric Bell state $\ket{\Psi^-}$ can be
achieved by projection onto the coherent state
$\ket{\alpha \ee^{-i\omega \tau}}$
only for those special initial conditions for which $\eta(\vec c,\phi)$ vanishes, such as  perfect resonant
interaction ($\delta=0$), 
equal initial weights  ($c_0=c_1$), and perfectly matched phases ($\phi=-\theta$).
A major challenge of our subsequent discussion will be the
development of a photonic measurement scheme by which such a perfect projection can be achieved
by this type of photonic postselection for all initial
conditions of the form of Eq. \eqref{initial}.
In the subsequent section 
it will be demonstrated that with the help of a Ramsey-type
interaction scenario which involves the two material 
qubits interacting with the modes of two different cavities
a material Bell state $\ket{\Psi^-}$ can be generated with almost unit fidelity
and success probability $|c_-|^2$ (see Eq. \eqref{initial}).

\subsection{Photonic postselection by balanced homodyne detection}
\label{phasespace}

Postselective projection of the tripartite quantum state 
$\ket{\Psi(\tau)}$ of 
Eq. \eqref{psi}  onto the 
coherent state $\ket{\alpha e^{-i\omega\tau}}$ 
can be achieved in a convenient way
with the help of
balanced homodyne detection.
As discussed in more detail in Appendix B
in a typical balanced homodyne detection measurement \cite{Lvovsky} 
the single-mode field state to be measured
is superposed coherently with an intense  coherent state
$\ket{|\alpha_{L}|e^{i\theta_L}}$ of a local oscillator by a $50\%$
reflecting beam splitter and the difference of photon numbers $n_-$
of the two modes emerging from the beam splitter is measured.
If the mode to be measured
is prepared in the quantum state $\hat{\rho}_F$ \cite{footnote},
the local oscillator
state is intense, i.e. $\mid \alpha_L\mid \gg 1$, and the homodyne
detection is performed with unit quantum efficiency, the detection scheme is equivalent to
a projective von Neumann measurement. In particular,
the probability of detecting
a difference photon number $n_-$ is given by
\begin{eqnarray}
P_{\theta_L}\left(
\tfrac{n_-}{\sqrt2|\alpha_L|}
\right) &=&
{\rm Tr}
\{
\hat{\rho}_F \ket{q_{\theta_L}} \langle q_{\theta_L}|
\}
\label{homodyne}
\end{eqnarray}
with the quadrature eigenstate $\ket{q_{\theta_L}}$ being determined
by the eigenvalue equation
\begin{eqnarray}
  \tfrac{1}{\sqrt2} \left(\hat{a}e^{-i\theta_L} + \hat{a}^{\dagger}e^{i\theta_L}\right)
\ket{q_{\theta_L}} &=& q_{\theta_L}\ket{q_{\theta_L}}
\label{quadraturestates}
\end{eqnarray}
with the  eigenvalues $q_{\theta_L}\in {\mathbb R}$ 
and with $a$ ($a^\dagger$) denoting 
the annihilation (creation) operator of the mode
to be measured.
Thus, in this limit a homodyne detection measurement is a von Neumann
measurement determined by the continuous set of orthonormal projectors
$\hat\pi(q_{\theta_L}) = \ket{q_{\theta_L}} \langle q_{\theta_L} |$.
This implies that a
postselective photonic measurement  with the phase 
$\theta_L = \phi -\omega\tau$ in an interval
$q_{\theta_L} \in 
(\sqrt2|\alpha|-\delta_L,\sqrt2|\alpha|+\delta_L)$ 
projects the field state
$\hat{\rho}_F=
{\rm Tr}_{A,B}\{ |\Psi(\tau)\rangle \langle \Psi(\tau)| \}$ 
onto the 
coherent state $\ket{\alpha e^{-i\omega\tau}}$ ($\alpha=|\alpha|e^{i \phi}$) with almost unit probability
provided the interval $\delta_L$ is chosen sufficiently large (compare with Eq. \eqref{error}
and the estimates
of Appendix B). 

\begin{figure}
  \includegraphics[width=.49\textwidth]{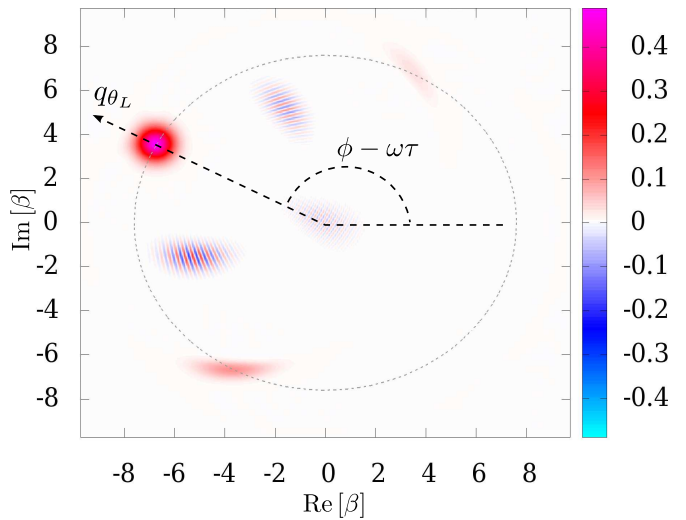}
\caption{\label{Wigner}
Wigner phase space distribution of the 
photonic state $\hat{\rho}_F$:
The  material state $\ket{\Psi^-}$ is solely paired with the
Gaussian peak. The rest of the states in Eq. \eqref{psi}
have contribution of  the three peaks and this explains the interference
fringes. $q_{\theta_{L}}$ represents the quadrature of a balanced homodyne
measurement. The interaction time is given by $\tau=\tau_{r}/4$ with the revival time $\tau_r$ of Eq. \eqref{revcoltime},  
$\omega=8 m\pi/\tau_r$ ($m\in\mathbb{N}_0$)
and the rest parameters correspond to those of Fig. \ref{overlap1}. 
}
\end{figure}

The Wigner phase space distribution  is a convenient way to 
visualize the field state (compare with Eq.  \eqref{psi})
\begin{equation}
  \hat{\rho}_F =|c_-|^2 \ket{\alpha  \ee^{-i \omega \tau}}\bra{\alpha  \ee^{-i \omega \tau}} 
 +\sum_{j=-1}^1\ket{\chi_j(\tau)}\bra{\chi_j(\tau)} 
\label{rhofieldstate}
\end{equation}
emerging from
the interaction between the two material quantum systems and 
the single-mode of the radiation field.
It is defined by \cite{Risken}
\begin{equation}
W(\beta,\beta^*) =
\frac{1}{\pi^2} \int {\rm Tr}\left\{\hat{\rho}_F \, 
\ee^{\zeta \hat{a}^\dagger-\zeta^* \hat{a}}\right\}
 \ee^{\beta\zeta^*-\beta^*\zeta} d^2\zeta
\label{Wignerf}
\end{equation}
with $\beta,\zeta \in {\mathbb C}$.
According to Eq. \eqref{initial} initially, 
i.e. at $\tau =0$, $\hat{\rho}_F=\ket{\alpha}\bra{\alpha}$ is a coherent state so 
that its
Wigner phase space distribution  is given by a Gaussian distribution. 
For $\tau > 0$ the mixed field state $\hat{\rho}_F$ always 
contains an admixture of the coherent state
$\ket{\alpha \ee^{-i\omega \tau}}$ which is strictly
correlated with the  material Bell state $\ket{\Psi^-}$ of the two qubits. 
The free time evolution of this coherent state reflects
the fact that the Bell state $\ket{\Psi^-}$ 
is not coupled to the single-mode radiation field.
However, due to the fact that in general $\ket{\Psi(\tau)}$ 
of Eq. \eqref{psi} is a tripartite
entangled state the Wigner distribution of $\hat{\rho}_F$
contains also additional maxima with interference
fringes in between. This is apparent from Fig. \ref{Wigner}. 
These interference fringes reflect the fact
that the freely evolving coherent state field state 
$\ket{\alpha \ee^{-i\omega\tau}}$ has finite overlaps with
the other field states $\ket{\chi_j(\tau)}~~(j=0,\pm1)$ 
constituting the mixed state $\hat{\rho}_F$.
To ensure that the interval $(\sqrt2|\alpha|-\delta_L,\sqrt2|\alpha|+\delta_L)$ of the homodyne measurement
does not include the interference fringes the
inequality $\delta_L<\tfrac{|\alpha|}{\sqrt2}\sin^2(\tfrac{\pi\tau}{\tau_r})$
has to be fulfilled for the interaction time $\tau$. 
This inequality can be derived from the coherent state approximation
by realising that the interference fringes have a Gaussian envelope and 
they are centered at $\sqrt2|\alpha|\cos^2(\tfrac{\pi\tau}{\tau_r})$ in $q_{\theta_L}$.
In Fig. \ref{Wigner} we used a detection time $\tau=\tau_r/4$ and $|\alpha|=7.6$
giving rise to the inequality $\delta_L<2.68701$, which still allows a very good
probability of projecting onto the desired coherent state (compare with
Eq. \eqref{error}). 

Finally, we would like to comment that
in a recent study by Rodrigues 
et al. \cite{Rodrigues} a similar protocol was introduced for the postselective preparation 
of a maximally entangled state by balanced homodyne photodetection when both material qubits
are prepared in the ground state.
In their scheme the resulting entangled state
has the inconvenience of having a time dependent relative phase.
In contrast, the method presented here can produce a perfect Bell state 
for certain initial conditions. In the following sections we will show how to
enlarge the class of initial conditions such that our method can be 
 extended to implement quantum teleportation and entanglement swapping protocols.

\section{A Ramsey-type photonic postselection scheme}
\label{Filter}
In this section
a generalization of the photonic postselection scheme of the previous section
is discussed which involves
a Ramsey-type matter-field interaction scenario with two cavities.
Ideally it allows the probabilistic
postselection of a two-qubit Bell state
with unit fidelity 
for arbitrary initial conditions of the material qubits. This photonic
postselection is achieved by
projection onto a coherent state which may be achieved with the help of
balanced homodyne photodetection. The success probability
of this postselective Bell-state projection
is determined by the initial condition of the material qubits.

\begin{figure}[t]
  \includegraphics[width=.49\textwidth]{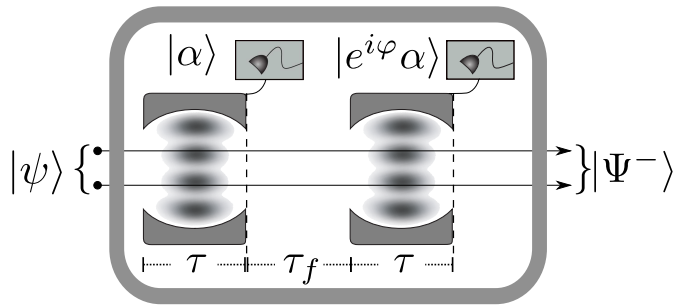}
  \caption{
  \label{filterfig}
  Ramsey-type interaction scenario for probabilistic  
  postselection of a two-qubit Bell state $\ket{\Psi^-}$:
 In a first step two qubits interact for a time $\tau$ with a 
  single photonic mode inside
  a cavity initially prepared in the coherent state $\ket\alpha$. 
Immediately afterwards 
  the resulting photonic state is projected onto 
the freely evolved coherent 
  state $\ket{\alpha \ee^{-i\omega\tau}}$.
  During the second step
  the two material qubits evolve freely for a time $\tau_f$. 
In the third step the two qubits
  interact with a second cavity initially prepared in a coherent state 
$\ket{\ee^{i\varphi}\alpha}$.
  At time $2\tau + \tau_f$ the photonic quantum state 
inside the second cavity
  is projected onto the freely evolved coherent state
  $\ket{\alpha \ee^{i\left(\varphi-\omega(2\tau+\tau_f)\right)}}$. Both photonic projections
can be achieved by homodyne detections which are depicted as detector inside
boxes.
  The two-qubit state resulting from this postselection process
  is the maximally entangled Bell state $\ket{\Psi^-}$. 
Ideally it is prepared with unit
  fidelity for arbitrary initial conditions of the two material qubits 
and with the success probability
  $|c_-|^2$.
  }
\end{figure}

Let us consider an interaction scenario as schematically depicted in 
Fig. \ref{filterfig}.
In a first step
two qubits 
interact with a single mode of the radiation field inside 
a cavity for a time $\tau$ so that
their interaction can be described by the Tavis-Cummings Hamiltonian
of Eq. \eqref{Hamilton}.
At time $\tau$ the resulting
tripartite qubit-field state is given in Eq. \eqref{psi} if initially
the radiation field is prepared in the coherent state $\ket{\alpha}$.
In the approximation of large mean photon numbers, i.e.
$\overline{n} \gg 1$,
and for interaction times $\tau$ in the plateau region of 
Fig. \ref{overlap1},
i.e. $\tau_c \ll \tau \ll \tau_r$,
projection of the resulting tripartite qubit-field state 
onto the freely evolved
coherent state $\ket{\alpha \ee^{-i\omega \tau}}$
yields the  two-qubit state of Eq. \eqref{projection} 
which reduces to
the maximally entangled Bell state provided the parameter 
$\eta(\vec c,\phi)$ vanishes.
However, in general a vanishing value of $\eta(\vec c,\phi)$
 can only be achieved for particular initially
prepared two-qubit states.

In order to achieve a vanishing value of $\eta(\vec c,\phi)$
for arbitrary initial conditions of the two-qubit system
a second identical interaction 
is enforced with a second cavity for a time $\tau$
(with $\tau_c \ll \tau \ll \tau_r$)
after a free time evolution of the two-qubit system
for a time $\tau_f$. The single mode of the second cavity 
interacting almost resonantly
with the two-qubit system is initially prepared in the coherent state
$\ket{\alpha \ee^{i\varphi}}$ which differs  
by a phase 
$\varphi$ from the initially prepared coherent field state $\ket{\alpha}$
of the first cavity. 
The intermediate free evolution of the two-qubit system during the second
step of this process
is governed by the free two-qubit Hamiltonian
$\hat H_{\rm a}=\hbar\omega_{\rm a}/2(\hat\sigma_A^z+\hat\sigma_B^z)$. This 
Hamiltonian 
only affects the phases accumulated  by the two-qubit states 
$\ket{0,0}$ and $\ket{1,1}$
appearing in Eq. \eqref{projection}. 
Thus, after the first photonic postselection
at time $\tau + \tau_f$ the tripartite state
involving the two material quantum systems and the mode of the second cavity
is given by
\begin{align}
  \ket{\Psi_1}=
  \left(\tfrac{c_-}{\sqrt P}\ket{\Psi^-}+
  \tfrac{
  \eta (\vec c,\phi)
  s}{\sqrt P} 
  \ket{\psi_{\phi-\omega_a\tau_f}}\right)
\ket{\alpha \ee^{i\tilde\varphi}}
\label{initial2}
\end{align}
with the success probability $P$ of Eq. \eqref{Prob} and the state 
  $\ket{\psi_{\phi-\omega_a\tau_f}}$ defined in
Eq. \eqref{undestate}.
The phase 
\begin{equation}
  \tilde\varphi=\varphi-\omega(\tau+\tau_f)
  \label{tildevarphi}
\end{equation}
takes into account the free evolution of the coherent state in the second cavity
which is assumed to be identical to the first cavity.
Immediately after the three-step Ramsey-type interaction sequence, i.e. 
at time $2\tau + \tau_f$, the resulting
two-qubit-field state is projected onto the freely evolved 
coherent state of the second cavity 
$\ket{\alpha \ee^{i{(\tilde\varphi-\omega\tau)}}}$.  
This projection can be evaluated in an analogous way as in the first projected state of
Eq. \eqref{projection}
yielding the postselected two-qubit quantum state 
\begin{align}
  \label{2ndint}
  \bra{\alpha \ee^{i{(\tilde\varphi-\omega\tau)}}}
  \ee^{-i\frac{\hat H'}{\hbar}\tau}
  \ket{\Psi_1}=&
  \tfrac{c_-}{\sqrt P}\ket{\Psi^-}+
  \\&
  \eta(\vec d,\phi+\tilde\varphi)
\tfrac{\eta(\vec c,\phi)
s^2
}{\sqrt P}
 \ket{\psi_{\phi+\tilde\varphi}}.
  \nonumber
\end{align}
The entries of
$\vec d=\left(\delta/\omega_{\overline n},
d_0
,-d_0^\ast
\right)$ 
represent the initial conditions of the state 
in Eq. \eqref{initial2} and according to the definition in Eq. \eqref{undestate} 
we get the value
$d_0=\ee^{-i(\theta-(\omega+\varpi_0)\tau+\phi-\omega_{\rm a}\tau_f)}/\sqrt2$.
They have to be inserted into Eq. \eqref{etas} in order to obtain
explicitly
\begin{align}
  \eta(\vec d,\phi+\tilde\varphi)=
  \tfrac{\omega_{\overline n}^2}{\Omega_{\overline n}^2}
  \left(
  \tfrac{\delta^2}{\omega_{\overline n}^2}
  +
 \cos{\Big(
  \tilde\varphi+(\omega+\varpi_0)\tau+\omega_{\rm a}\tau_f
  \Big)}
  \right).
\end{align}
The Hamiltonian $\hat H'$ in Eq. \eqref{2ndint} has the same form as Eq. \eqref{Hamilton}
and we use the primed notation to distinguish the mode of the second cavity
from the mode of the first cavity.
The parameter $\eta(\vec d,\phi+\tilde\varphi)$ can vanish if the  
initial phase of the second coherent state $\varphi$  
is chosen in such a way that the conditions 
\begin{align}
  \varphi=
  \arccos{\left(-\frac{\delta^2}{\omega_{\overline n}^2}\right)}
  -\varpi_0\tau
  -\delta\tau_f,
  \label{condition}
\end{align}
and 
$\delta\le\omega_{\overline n}= g\sqrt{4\overline n-2}$
are fulfilled. In the case of perfect resonance 
($\delta=0$) $\varphi$ takes  the value of $\pi/2$.
If the condition of Eq. \eqref{condition} is fulfilled 
the projection onto the coherent state $\ket{\alpha e^{i(\tilde\varphi-\omega\tau)}}$
postselects the Bell state $\ket{\Psi^-}$ and this occurs with a probability
of $P'=|c_-|^2/P$ (compare with Eq. \eqref{2ndint}).
Because both projections are independent
the overall success probability of this scheme is given by
\begin{align}
  P_T=P P' =|c_-|^2.
  \label{}
\end{align}

Both projections onto the relevant
coherent states of the single-mode radiation fields  
can be achieved by balanced homodyne detection of the relevant photons
by appropriate choices of the phases of the local oscillators.
For the homodyne measurement at time $\tau$ 
one has to choose $\theta_{L} = \phi- \omega \tau$ and for the
corresponding homodyne detection at time $2\tau + \tau_f$ the phase
of the local oscillator has to adjusted to the value 
$\theta_{L}' = \phi + \varphi - \omega (2\tau+\tau_f)$. Remarkably,
this probabilistic postselective preparation
of the two-qubit Bell state $\ket{\Psi^-}$ can be achieved
for arbitrary initially prepared quantum states of the 
two material quantum systems. Ideally this preparation can be realized 
with unit fidelity and with a success probability $|c_-|^2$ 
which depends on the
initially prepared two-qubit state.

\begin{figure}
  \includegraphics[width=.49\textwidth]{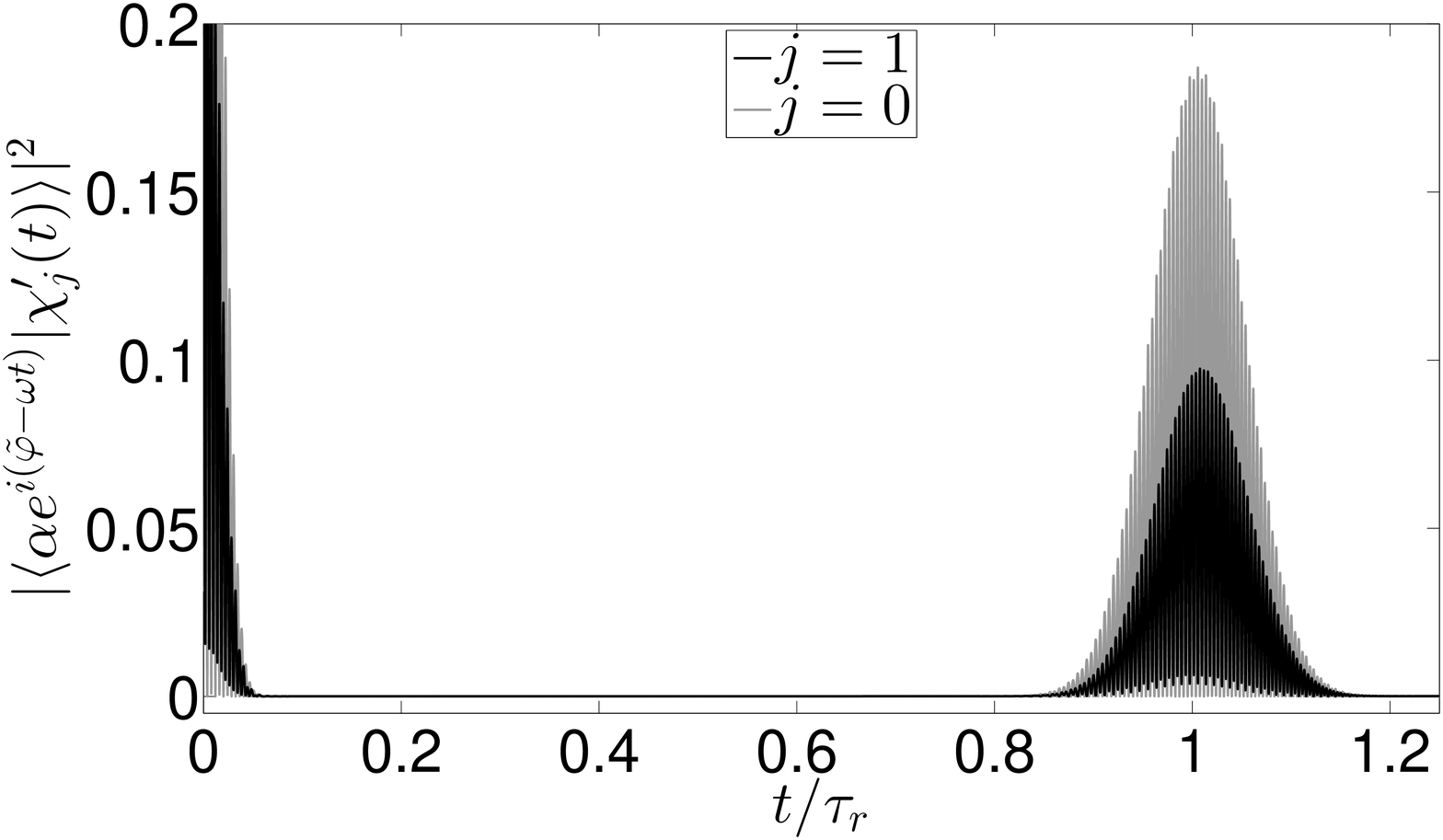}
  \caption{\label{overlap2}Overlap between the exact photonic states 
$\ket{\chi_j'(t)}$ of the second cavity and
  the coherent state $\ket{\alpha \ee^{i(\tilde\varphi-\omega t)}}$ 
characterizing the third step of the
  Ramsey-type postselection scheme. 
  The initial material state is taken from Eq. \eqref{projection} for an interaction
  time of $\tau=\tau_r/4$. The parameters $\alpha$, $c_-$, $c_1$, $c_+$, $c_0$, and $\delta/g$
 are set to the same value as in Fig. \ref{overlap1}. All three overlaps vanish in 
 the plateau region.  
}
\end{figure}

\begin{figure}
  \includegraphics[width=.49\textwidth]{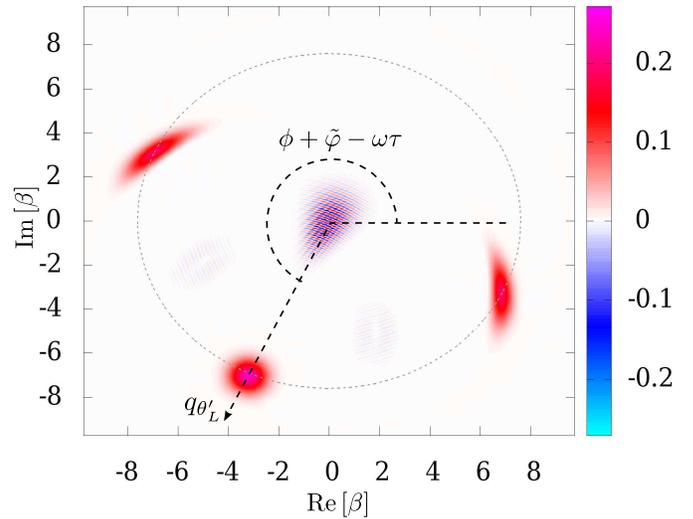}
\caption{\label{Wigner2}Wigner phase space distribution of the  
photonic quantum state 
in the second cavity after the third step of the Ramsey-type 
postselection scenario:
The Gaussian peak is correlated with the  material state $\ket{\Psi^-}$.
The rest of the photonic states in Eq. \eqref{psi}
have no contribution of the Gaussian peak. This explains the vanishing
interference fringes with the other two peaks. $q_{\theta_{L}'}$ 
represents the quadrature of a balanced homodyne
measurement.
The interaction time is given by $\tau=\tau_{r}/4$ with the revival time $\tau_r$ of Eq. \eqref{revcoltime},  
$\omega=8 m\pi/\tau_r$ ($m\in\mathbb{N}_0$). 
The other parameters correspond to those of Fig. \ref{overlap2}. 
}
\end{figure}

In Fig. \ref{overlap2} the overlaps between the photonic field states
$\ket{\chi_j(t)}$ and the freely evolved coherent state 
$\ket{\alpha \ee^{i(\tilde\varphi-\omega\tau)}}$ are shown for the initial condition of Eq. \eqref{initial2}.
The overlaps clearly vanish for times $\tau$ in the plateau region
i.e. $\tau_c \ll  \tau \ll \tau_r$.
The corresponding Wigner function of the field state is depicted in
Fig. \ref{Wigner2}. 
Here, consistent with these vanishing overlaps
the interference fringes between the freely evolving field state and the residual field states are not present. 
This demonstrates that the coherent state 
$\ket{\alpha \ee^{i(\tilde\varphi-\omega\tau)}}$ is solely paired with
the Bell state $\ket{\Psi^-}$ which 
can be prepared with unit fidelity by photonic postselection.

It is worth mentioning that this probabilistic preparation of 
the two-qubit
Bell state $\ket{\Psi^-}$ by two time-delayed homodyne measurements 
also works
in more general situations which involve 
different detunings and different
dipole coupling phases in both cavities, for example.
In such cases one would have to add to Eq. \eqref{condition} 
the difference
between both of the dipole coupling phases, i.e. $\theta-\theta'$,
the detuning between cavities times the interaction time 
$(\omega'-\omega)\tau$,
and to perform the replacements
$\delta^2/\omega_{\overline n}^2\to \delta\delta'/\omega_{\overline n}
\omega_{\overline n'}$ and $\delta\tau_f\to\delta'\tau_f$.


\section{Different qubits}
\label{stability}
In this section we explore the case of  different
coupling strengths of the qubits to the field as well as
different  transition frequencies. 
This is of interest for any 
experimental realization of the proposed
scheme. To this end we choose
to define the coupling strength of qubit $A$ ($B$) to the radiation field
as $g_A=g+\varepsilon_g$, ($g_B=g-\varepsilon_g$). The transition frequency of qubit $A$ ($B$) is detuned 
from the frequency of the cavity mode as described  by the equation
$\delta_A=\delta+\varepsilon_\delta$ ($\delta_B=\delta-\varepsilon_\delta$).

In this situation  the state $\ket{\Psi^-}\ket{n}$ is no longer an 
eigenstate of the Hamiltonian and therefore 
the time dependent state vector of the complete
system is given by
\begin{align}
  \ket{\Psi(t)}&=\ket{\Psi^-}\ket{\chi_2(t)}
  \label{psieps}
  \\&
  +\ket{1,1}\ket{\chi_{1}(t)}
  +\ket{\Psi^+}\ket{\chi_{0}(t)} 
  +\ket{0,0}\ket{\chi_{-1}(t)}.
  \nonumber 
\end{align}
In contrast to Eq. \eqref{psi} the photonic state 
$\ket{\chi_2(t)}$ is in general no longer a 
coherent state. 

\begin{figure}
  \includegraphics[width=.49\textwidth]{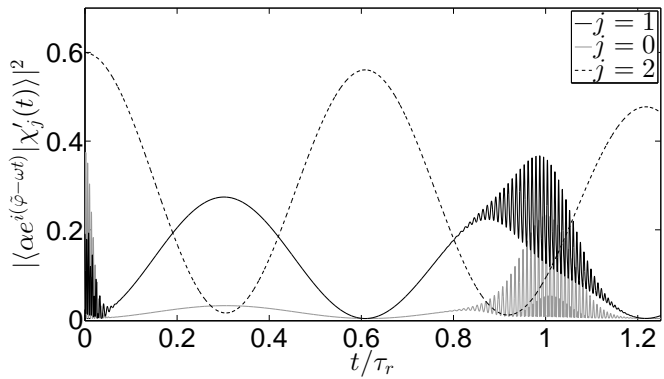}
  \caption{\label{overlapeps}Overlap between the exact photonic states 
$\ket{\chi_j'(t)}$ of the second cavity and
  the coherent state $\ket{\alpha \ee^{i(\tilde\varphi-\omega t)}}$ 
characterizing the third step of the
  Ramsey-type postselection scheme. 
  The initial material state is taken from Eq. \eqref{projection} for an interaction
  time of $\tau=\tau_r/4$. The parameters $\alpha$, $c_-$, $c_1$, $c_+$, $c_0$, and $\delta/g$
 are set to the same value as in Fig. \ref{overlap1}. In addition we consider an
 asymmetry in the coupling strengths of $\varepsilon_g/g=0.007$ but keep
 equal detunings, i.e. $\varepsilon_\delta=0$.
}
\end{figure}

In Fig. \ref{overlapeps} we present an exact numerical calculation
of the overlaps of the photonic field states of the second cavity
with the coherent state $\ket{\alpha \ee^{i(\tilde\varphi-\omega\tau)}}$.
This is the analog of Fig. \ref{overlap2} but with an asymmetry
in the coupling strengths of $\varepsilon_g/g=0.007$. 
In addition we have included the overlap with the state $\ket{\chi_2(t)}$
which is not a constant. 
One notes
the emergence of additional Rabi oscillations. In
Appendix \ref{DiffAt} we show that this Rabi frequency increases as a function of 
$\varepsilon_g$ and  $\varepsilon_\delta$. 
The oscillations are also damped and undergo the typical 
collapse and revival phenomena.

\begin{figure}
  \includegraphics[width=.49\textwidth]{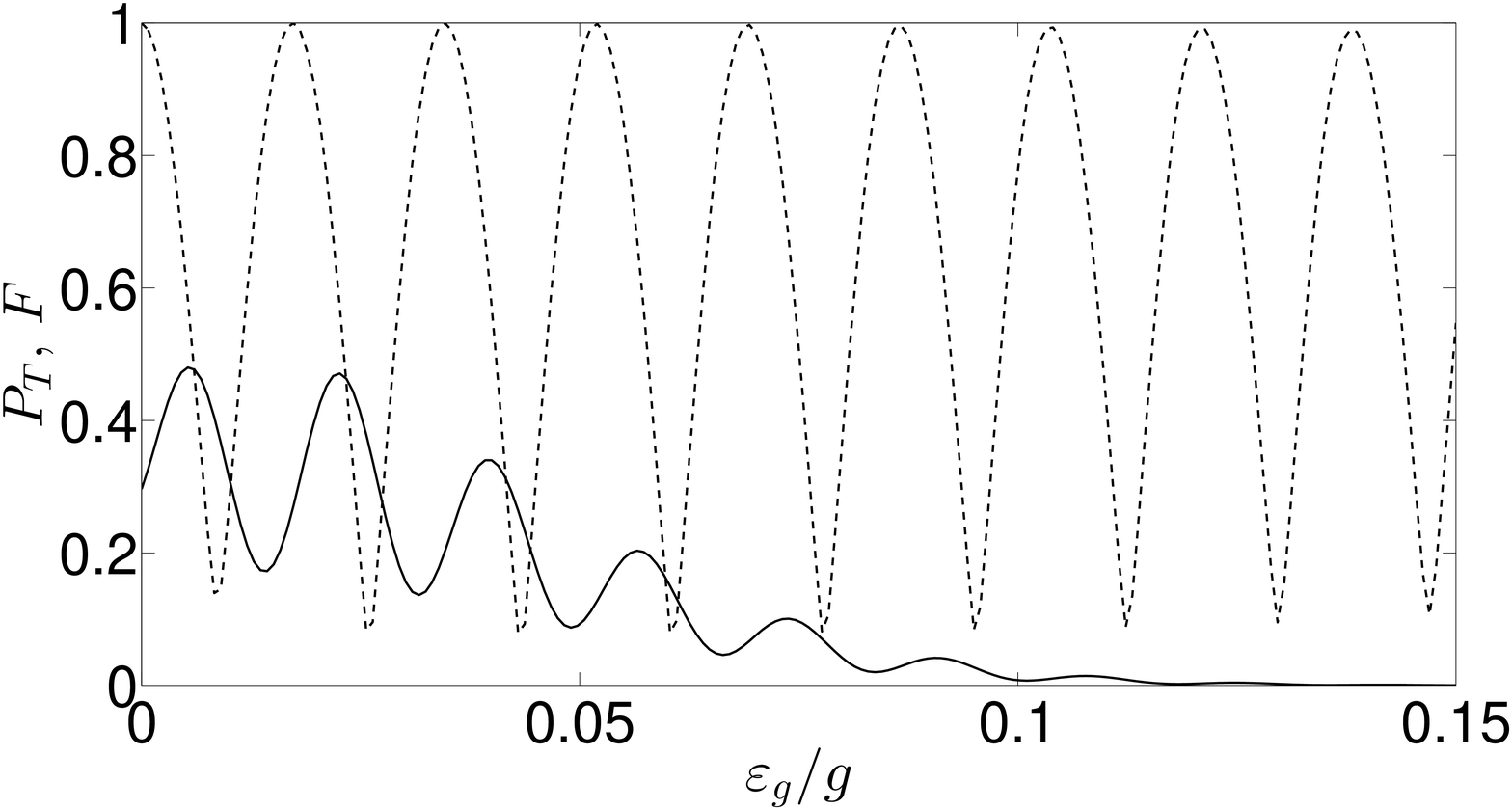}
  \caption{\label{fidsuc}
  Success probability $P_T$ (full line) and fidelity $F$ (dashed line)
  of achieving the state $\ket{\Psi^-}$ as a function of the
  asymmetry in the coupling strengths $\varepsilon_g$ and
  using the scheme of Fig. \ref{filterfig}.
  We take $\varepsilon_\delta=0$. The rest of the parameters and
  the initial conditions are the same of Fig. \ref{overlap1}. 
}
\end{figure}

To evaluate how unequal coupling strengths influence our scheme
presented in Sec. \ref{Filter} we evaluate
the overall success probability $P_T$ and the fidelity $F$ of achieving the Bell state
$\ket{\Psi^-}$. 
In Fig. \ref{fidsuc} we present an exact numerical calculation for both of these 
quantities as a function of the difference $\varepsilon_g$ between couplings strengths
of the qubits to the cavity mode. In this example we took 
equal detunings, i.e. $\varepsilon_\delta=0$. 
The fidelity displays an oscillatory behaviour and it attains
values close to unity in a periodic way. The success probability also oscillates 
and decays as a function of $\varepsilon_g$. Both effects are
consequences of the collapse and revival phenomena of the Rabi oscillations
induced by unequal couplings.

The frequency of the Rabi oscillations increases
as a function of $\varepsilon_g$ and the maxima of the fidelity occurs
at values of $\varepsilon_g$ where the Rabi oscillations complete a cycle
at interaction time $\tau=\tau_r/4$ (compare with Eq.(\ref{revcoltime}). We can
estimate that this happens for integer multiples of
$\varepsilon_g/g\to 4g^2/\omega_{\overline n}^2\approx 1/\overline n$
(compare with Eq.(\ref{frequenciesTaylor}). Similar behaviour
of the fidelity occurs for an asymmetry in the detunings so that the cycles are completed
at integer multiples of $\varepsilon_\delta/g\to4g/\delta$. We can conclude
that unequal coupling strengths between the qubits to the radiation field and unequal
detunings have to fulfill the requirements $\varepsilon_g/g\ll1/2\overline n$
and $\varepsilon_\delta/g \ll 2g/\delta$ because it is at these values
where the first minimum of the fidelity is attained. This means that the scheme is 
sensitive to variations of the coupling strengths but more robust with respect
to small variations of the detunings. In Appendix \ref{DiffAt} we show
details of the derivations of these conditions. 

A possible experimental realization of the Bell projection scheme involving nowadays 
technology could involve flying atoms and single mode cavities.
Different coupling strengths to the cavity mode can arise from 
the different paths on which the atoms cross the  electromagnetic field mode inside the cavity.
Therefore, if we consider two mirrors  of a cavity facing each other along the $z$ axis,
for example, a typical position dependent
coupling strength can be modelled by
\begin{equation}
 g(x,y,z)=g_0 \sin\left(\frac{2 \pi z}{\lambda}\right)\ee^{-\frac{x^2+y^2}{w^2}}
\end{equation}
with $\lambda$ and $w$
denoting the wavelength of the cavity and the mode waist. Thereby, 
the spatial positions of the flying atoms are chosen so that their $x$ and $y$
coordinates are the same and 
they pass through the cavity at different values of $z$.
In order to achieve strong coupling 
both atoms should be located at the antinodes of the radiation field.
However, even in this case unequal couplings to the field mode may result from
inaccuracies in the positions of the atomic paths.
As we know that
our scheme works for $(g_A-g_B)/(g_A+g_B)<1/2\bar{n}$ let us address the
question for which inaccuracies in the positions of the atomic paths
this condition can still be fulfilled. 

For this purpose let us consider the recent experiment of Ref.\cite{Gleyzes}
with flying Rydberg atoms.
In this experiment the mirrors are positioned at a distance of $2.7$ cm,
the cavity is resonant at $51.1$ GHz, the maximum 
coupling is given by $g_0/2\pi=51$ kHz and the waist is $w=6$ mm.
The experienced change in the coupling strength due to the waist is well under control
because the experimental study integrates 
the collected data over the flying time through the cavity.
We can now estimate the allowed deviations $\epsilon_{z,A}$, $\epsilon_{z,B}$ in the
positions of the atoms by
\begin{equation}
  \left|
 \frac{\sin\left(\frac{2 \pi (z_A+\epsilon_{z,A})}{\lambda}\right)-\sin\left(\frac{2 \pi (z_B-\epsilon_{z,B})}{\lambda}\right)}
 {\sin\left(\frac{2 \pi (z_A+\epsilon_{z,A})}{\lambda}\right)+\sin\left(\frac{2 \pi (z_B-\epsilon_{z,B})}{\lambda}\right)}\right|<1/2\bar{n}.
\end{equation}
Assuming that the ideal positions $z_A$ and $z_B$ are such that
$\sin\left(2 \pi z_A/\lambda\right)=\sin\left(2 \pi z_B/\lambda\right)=1$ and that
the deviations 
$\epsilon_{z,A}$, $\epsilon_{z,B}$ are below $1$ mm we obtain
\begin{equation}
|\epsilon_{z,A}-\epsilon_{z,B}|<\lambda/\pi \arctan\left(\frac{1}{2\bar{n}}\right).
\label{asscond}
\end{equation}
Thus, if the deviations are similar for both paths, i.e. $\epsilon_{z,A}=\epsilon_{z,B}$,
the above condition is always fulfilled. Otherwise, for
$\bar{n} \sim 10^{2}$ the difference of these deviations must obey the relation
$|\epsilon_{z,A}-\epsilon_{z,B}|<10 \mu$m. Therefore,
if the average number of photons $\overline{n}$
is not too large, current experiments are precise enough to realize
the condition of Eq. \eqref{asscond}.

\section{Quantum information processing}
\label{QSchemes}
In this section it is demonstrated how the Ramsey-type interaction scheme of 
Sec. \ref{Filter}
can be used for implementing probabilistic quantum teleportation and entanglement
swapping. Thereby, the crucial feature is exploited that 
ideally
this Ramsey-type interaction
scheme allows to postselect 
a Bell state of two material qubits
with unit fidelity for a large class of initial conditions
of the two material qubits. As this postselection
procedure can be implemented with the help of
balanced homodyne photodetection it offers
interesting perspectives for current applications 
in quantum information processing.

\subsection{Entanglement-assisted Teleportation}
\begin{figure}
  \includegraphics[width=.49\textwidth]{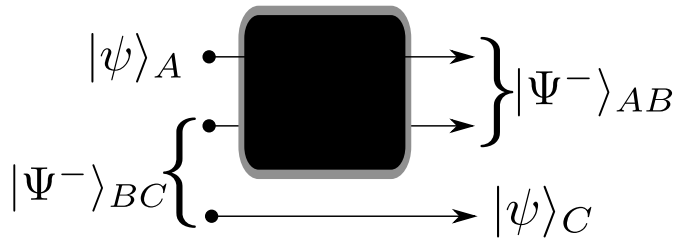}
\caption{
\label{filterfigtel}
A probabilistic quantum teleportation protocol
based on the Ramsey-type photonic postselection
scheme of the Bell state
$\ket{\Psi^-}$ as discussed in Sec. \ref{Filter} and depicted by Fig. \ref{filterfig}.
}
\end{figure}

The goal of entanglement-assisted quantum teleportation is to transfer the unknown
state of a quantum system, say $A$, to
another quantum system, say $C$.
So, let us consider  three material qubits $A$, $B$, and $C$ as depicted in Fig. \ref{filterfigtel}
with the qubit $B$ acting as an ancilla system.
Initially qubit $A$ is prepared in the unknown quantum state
\begin{align}
  \ket{\psi}_{A}=
  a\ket{0}_A + b\ket 1_A,\,|a|^2+|b^2|=1.
  \label{telstate}
\end{align}
Thus, in order to implement a photon-assisted quantum teleportation protocol
let us consider the initially prepared four-partite quantum state
\begin{align}
  \ket{\Psi_0^{\rm tel}}
  =&
  \ket{\psi}
  _A\otimes\ket{\Psi^-}_{BC}\otimes\ket\alpha,
\end{align}
which involves the three material qubits $A$, $B$, $C$ and the initially
prepared single-mode coherent quantum state $\ket{\alpha}$ 
of the radiation field.
This initially prepared four-partite quantum state can be represented
in the equivalent form
\begin{align}
  \ket{\Psi_0^{\rm tel}}&=
  -\tfrac{1}{2}\ket{\Psi^-}_{AB}
\ket{\psi}_C\ket\alpha
  \nonumber\\&
  +\left(\tfrac{b}{2}\ket{\Psi^+}_{AB}+\tfrac{a}{\sqrt2}\ket{00}_{AB}\right)
\ket{1}_C\ket \alpha
  \nonumber\\&
  -\left(
  \tfrac{b}{\sqrt2}\ket{11}_{AB}+
  \tfrac{a}{2}\ket{\Psi^+}_{AB}
  \right)
  \ket{0}_C\ket \alpha .
  \label{intel}
\end{align}
Furthermore, let us assume that
atoms $A$ and $B$  interact with the single mode of the radiation field
inside a cavity so that this interaction can be described by
the Hamilton operator of the Tavis-Cummings model of Eq. \eqref{Hamilton}.
As the Bell state $\ket{\psi^-}_{AB}$ is an invariant state of the Tavis-Cummings
model the photonic state evolves freely as a harmonically oscillating coherent state.
From  our discussion in Sec. \ref{Filter} it is known
that after an interaction time  $\tau$ with $\tau_c \ll \tau \ll \tau_r$ 
successful projection onto the photonic state $\ket{\alpha e^{-i\omega \tau}}$
results in the unnormalized  tripartite material quantum state
(see Eq. \eqref{projection})
\begin{align}
  \ket{\psi^{\rm tel}}
  &=
  \braket{\alpha \ee^{-i\omega t}}{\Psi^{\rm tel}(\tau)}=
  -\tfrac{1}{2}\ket{\Psi^-}_{AB}\ket{\psi}_{C}+
  \nonumber\\
  &+
  s
 \ket{\psi_{\phi}}_{AB}
  \left(
 \eta(\vec d_1,\phi)\ket 1_C+
 \eta(\vec d_0,\phi)\ket 0_C
  \right).
  \label{projtel}
\end{align}
Thereby, the state $\ket{\psi_{\phi}}_{AB}$ is given by Eqs. \eqref{projection}, \eqref{projection2}, \eqref{undestate} and
the amplitudes  in Eq. \eqref{projtel} are defined by the initial
conditions encoded in the vectors
$\vec d_1= (b/2,a/\sqrt{2},0)$ and $\vec d_0 = (-a/2,0,-b/\sqrt{2})$ 
according to the definition of $\eta(\vec c,\phi)$ in Eq. \eqref{etas}.

This projection onto the state \eqref{projtel} takes place with 
probability
\begin{align}
  P^{\rm tel}=\tfrac{1}{4}+|s|^2(
  |\eta(\vec d_0,\phi)|^2+|\eta(\vec d_1,\phi)|^2
  )
  \left(1+\tfrac{\delta^2}{\omega_{\overline n}^2}\right).
  \label{probtel}
\end{align}
Now, let us assume that
subsequently the quantum systems $A$ and $B$
interact for a time $\tau$
with a second cavity prepared in the single-mode coherent state 
$\ket{\alpha e^{i\varphi}}$. If the relative phase $\varphi$
fulfills
the condition of Eq. \eqref{condition} a second projection
onto the freely evolved coherent state $\ket{\alpha e^{i(\tilde\varphi-\omega\tau)}}$
results in the teleported quantum state
\begin{align}
  \ket{\Phi^{\rm tel}}=
  \ee^{i\pi}\ket{\Psi^-}_{AB}
  \otimes
  \ket{\psi}_C.
  \label{telfinal}
\end{align}
This second photonic projection takes place with probability 
${P^{\rm tel}}'=1/(4P^{\rm tel})$ so that 
the overall success probability of this entanglement-assisted
quantum teleportation protocol is independent 
of the initial conditions of the state to be teleported
 and is given by
\begin{align}
  P^{\rm tel}_T=0.25.
  \label{}
\end{align}

\subsection{Entanglement swapping}
\begin{figure}
  \includegraphics[width=.49\textwidth]{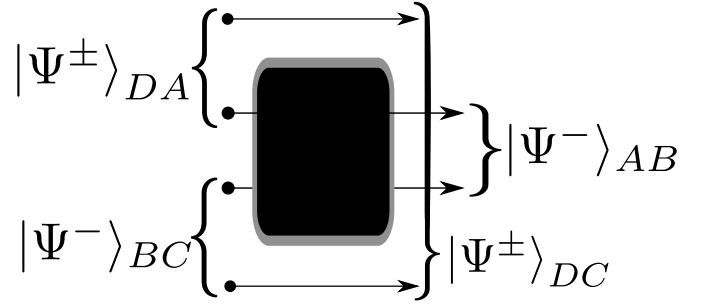}
  \caption{\label{filterfigswap}
A probabilistic
entanglement swapping protocol
based on the Ramsey-type photonic postselection
scheme of the Bell state
$\ket{\Psi^-}$ as discussed in Sec. \ref{Filter} and depicted by Fig. \ref{filterfig}.
}
\end{figure}

A major aim of entanglement swapping is to produce entanglement
between two distant quantum systems, say $C$ and $D$, with the help of two 
uncorrelated pairs
of entangled quantum systems, 
say $AD$ and $BC$. 
Let us consider  four material qubits $A$, $B$, $C$, and $D$ 
as depicted in Fig. \ref{filterfigswap}. Initially, the qubit pairs $BC$ and $AD$ 
are prepared in maximally entangled Bell states and an ancillary photonic field
mode is prepared in a coherent state $\ket{\alpha}$ so that the five-partite
initially prepared quantum state is given by
\begin{align}
&\ket{\Psi_0^{\rm sw}}=
\ket{\Psi^\pm}_{DA}
\ket{\Psi^-}_{ BC}
\ket\alpha.
\end{align}
This initial state can be represented in the equivalent form
\begin{align}
\label{initialswap}
\ket{\Psi_0^{\rm sw}}
&=
-\tfrac{1}{2}
\ket{\Psi^-}_{AB}\ket{\Psi^\pm}_{DC}\ket\alpha
-\tfrac{1}{2}\ket{1,1}_{AB}\ket{0,0}_{DC}\ket\alpha
\nonumber\\&
+\tfrac{1}{2}\ket{\Psi^+}_{AB}\ket{\Psi^\mp}_{DC}\ket\alpha
\pm\tfrac{1}{2}\ket{0,0}_{AB}\ket{1,1}_{DC}\ket\alpha.
\end{align}

A 
Bell projection on qubits $A$ and $B$ 
is capable of swapping
entanglement to qubits $C$ and $D$. 
For this purpose  qubits
$A$ and $B$ interact with the ancillary photonic field mode inside a
cavity for a time $\tau$ with
$\tau_c \ll \tau\ll\tau_r$. If this interaction can be described by the Tavis-Cummings
Hamiltonian of Eq.\eqref{Hamilton} we can take advantage from the fact that the two-qubit
Bell state $\ket{\Psi^-}_{AB}$ is an invariant quantum state under the Hamiltonian
of Eq.\eqref{Hamilton} so that the photonic field state it is correlated with 
according to
Eq.\eqref{initialswap} evolves freely as an oscillating coherent state. Thus,
projection of the five-partite quantum state $\ket{\Psi(\tau)}$
onto the coherent state $\ket{\alpha  e^{-i\omega \tau}}$ yields
the unnormalized four-partite qubit state (see Eq. \eqref{projection})
\begin{align}
  &\ket{\psi^{\rm sw}}=\braket{\alpha \ee^{-i\omega \tau}}{\Psi^{\rm sw}(\tau)}=
  -\frac{1}{2}\ket{\Psi^-}_{AB}\ket{\Psi^\pm}_{DC}+
  \nonumber\\
  &s\ket{\psi_{\phi}}_{AB}
  \left(
  \eta(\vec d_{00},\phi)\ket{0,0}_{DC}+ 
  \eta(\vec d_+,\phi)\ket{\Psi^\mp}_{DC}+
  \right.
  \nonumber\\
  &\left.
  \qquad
  \qquad
  \eta(\vec d_{11},\phi)\ket{1,1}_{DC}
  \right)
  \label{projsw}
\end{align}
with the initial conditions of Eq. \eqref{initialswap}
represented by the vectors 
$\vec d_{00} = (0, 0,-1/2)$, 
$\vec d_{+} = (1/2,0, 0)$
and $\vec d_{11} = (0,\pm 1/2,0)$.
These initial conditions have to be
substituted into the definition
of $\eta(\vec c,\phi)$ in Eq. \eqref{etas}.
The success probability of this photonic projection  is given by
\begin{align}
  P^{\rm sw}&=\tfrac{1}{4}+|s|^2
  \left(1+\tfrac{\delta^2}{\omega_{\overline n}^2}\right)\times
  \nonumber\\
  &\left(
  |\eta(\vec d_{+},\phi)|^2+|\eta(\vec d_{00},\phi)|^2+
  |\eta(\vec d_{11},\phi)|^2
  \right)
  .
  \label{}
\end{align}
In order to achieve projection onto the Bell state 
$\ket{\Psi^-}_{AB}$ with unit fidelity,
qubits $A$ and $B$ interact with a second single mode of the radiation field
inside a second cavity for a time $\tau$. Thereby, the radiation field
is prepared in a coherent state $\ket{\alpha e^{i\varphi}}$
so that condition \eqref{condition} is fulfilled. 
According to our discussion in section \ref{Filter} 
after the projection onto the second coherent state
$\ket{\alpha e^{i(\tilde\varphi-\omega \tau)}}$ the final four-partite qubit state
is given by
\begin{align}
  \ket{\Phi^{\rm sw}}=\ee^{i\pi}\ket{\Psi^-}_{ AB}\ket{\Psi^\pm}_{DC}.
  \label{finalsw}
\end{align}
This second photonic projection is achieved with a success probability of 
$1/(4P^{\rm sw})$. 
Multiplying the probabilities of both
photonic projections yields the overall success probability of this probabilistic
entanglement swapping procedure, namely
\begin{align}
  P_T^{\rm sw}=0.25.
  \label{}
\end{align}

Let us finally address the question to which extent the entanglement
swapping procedure discussed here may offer interesting perspectives
for current experimental activities in realizing a quantum repeater.
The experiments of  Gleyzes {\it et al}.
\cite{Gleyzes} have demonstrated that controlled interaction between
Rydberg atoms crossing several 
cavities and interacting with single modes of the radiation field prepared inside these 
cavities is possible.  Thus, the entanglement swapping protocol discussed here may be
integrated in a hybrid quantum repeater setup as proposed by van Loock et. al. \cite{vanLoock} or
in a setup based on almost resonant matter-field interaction \cite{Bernad},
for example, in the following way.
In a first step entanglement is generated between neighboring
stations by passing material qubits through different cavities. 
Due to  lossy transmission channels between the stations
entanglement purification \cite{Gonta} may be performed.
In a second step the previously discussed
entanglement swapping procedure is applied at each station.
Even if the qubits $A$ and $B$ are destroyed after the entanglement swapping 
procedure of Fig. \ref{filterfigswap} qubits $D$ and $C$ are still prepared 
in a Bell state $\ket{\Psi^\pm}_{DC}$.
Problems arising from
the fact that 
radiatively long lived stable electronic levels should be used
as 
material qubits may be resolved with the help of appropriately applied $\pi$-pulses. They
transform radiatively long lived electronic states to higher electronic levels which can
be excited almost resonantly by photons easily. 
Furthermore, recent experiments indicate that also the
condition of negligible spontaneous photon emission
into other modes of the radiation field during
the interaction between the qubits and the almost resonantly
coupled cavity modes can be fulfilled.
Although the direct experimental investigation of the two-qubit Tavis-Cummings model by
Casabone et al. \cite{Casabone}, for example, performed on
trapped $~^{40}\text{Ca}^+$ ions reports a ratio between  the vacuum Rabi frequency $g$ and the
spontaneous decay rate $\Gamma$ of the qubits as small as $g/\Gamma=0.68$, 
the experiment
of Colombe et al. \cite{Colombe} 
reports significantly higher ratios as large as $g/\Gamma=71.66$.
Thus, the experimental realization of the
dynamical regime of negligible spontaneous photon emission
into other modes of the radiation field
is within reach of nowadays experimental possibilities.
 
\section{Conclusions}
\label{Conclusions}

We have discussed a quantum electrodynamical implementation
of a probabilistic Bell measurement capable of projecting an
arbitrary initial state of two material qubits perfectly
onto a Bell state with success probability given by the initial probability
weight of this Bell state.
It has been demonstrated how this Bell measurement can be used as a building
block for implementations of entanglement-assisted teleportation
and entanglement swapping protocols both of which can be achieved
with almost unit fidelity and $25\%$ success probability.
This Bell measurement is
performed by entangling the two material qubits to be measured with
single modes of the radiation field in a Ramsey-type interaction sequence
and postselecting the resulting photon fields with the help of  balanced
homodyne photodetection. Within the dipole- and rotating wave approximation
the almost resonant quantum electrodynamical
matter-photon interaction involved in this Bell measurement can be described by the
two-qubit Tavis-Cummings model. The protocols presented
take advantage of a characteristic
feature of this particular interaction model, namely the existence of
an invariant two-qubit Bell state which does not couple to the photons.
Therefore, if initially the ancillary photon fields are prepared in coherent states this
invariant Bell state will always remain correlated with these
coherent states which evolve freely despite the presence of the
quantum electrodynamical matter-photon coupling. If the interaction times
of the Ramsey-type interaction sequence and the initial phases of the
coherent photon states are chosen appropriately ideally these coherent states can be
distinguished perfectly from the residual photon states which are correlated with
the other components of the material two-qubit quantum state. 
This offers the possibility to postselect these
coherent components of the photon 
state by balanced homodyne photodetection thus preparing
a perfect material two-qubit Bell state
with unit fidelity independently of the two-qubit state which has been prepared before
the interaction with the radiation field.
The properly chosen interaction times and phases of the coherent photon states involved
in this Ramsey-type interaction sequence exploit characteristic dynamical properties
of the collapse and revival phenomena of the Tavis-Cummings model and ensure that
this postselective unit-fidelity Bell state projection
can be achieved. 
It is this latter property which enables
the use of this probabilistic Bell measurement
as a basic building block for probabilistic
entanglement-assisted quantum teleportation. Furthermore, this probabilistic
Bell measurement may also be used for implementing entanglement swapping and may thus be of
particular interest for current experimental efforts aiming at the
realization of hybrid quantum repeaters.

In view of significant recent progress in quantum state engineering
and in the distribution of remote entanglement the postselective Bell measurement,
the quantum teleportation and entanglement swapping protocols discussed here
may offer interesting perspectives for future applications. 
Possible applications may not only include
quantum optical implementations of hybrid quantum repeaters
and quantum communication networks but also condensed-matter implementations of qubits which are almost resonantly
coupled to coherent states of microwave fields.

\begin{acknowledgments}
This work is supported by the BMBF project Q.com.
\end{acknowledgments}

\appendix
\section{Time evolution of the almost resonant two-qubit Tavis-Cummings model}
\label{Appsol}
In this section the time evolution of the two-qubit Tavis-Cummings model is discussed.
Let us consider the situation of an almost resonant coupling between
the qubits and the single mode of
the cavity field. For simplicity we omit the labels
of the qubits while taking the convention of keeping the order $A,B$, i.e. 
$\ket{1}_A\ket{0}_B=\ket{1,0}$. 
It is apparent that the state $\ket{\Psi^-}\ket{n-1}$ is an eigenstate of the 
Hamiltonian in Eq. \eqref{Hamilton}. 
Furthermore, 
the number of excitations of the two-qubit-field system
$\hat a^\dagger \hat a+\tfrac{1}{2}(\hat \sigma_A^z+\hat\sigma_B^z)$ 
is
a constant of motion of
the Hamiltonian \eqref{Hamilton} of the Tavis-Cummings model. 
This number of excitations is diagonal in the basis
\begin{align}
  &\{\ket{\Psi^-}\ket n\}_{n=0}^\infty\oplus
  \{\ket{0,0}\ket0\}\oplus
  \nonumber\\&
  \{\ket{\Psi^+}\ket{0},\ket{0,0}\ket{1}\}\oplus
  \nonumber\\&
  \{\ket{1,1}\ket{n-2},
  \ket{\Psi^+}\ket{n-1}
  ,\ket{0,0}\ket{n}\}_{n=2}^\infty,
  \label{basisAB}
\end{align}
and has a $3$-fold degenerate spectrum for any fixed value of
$n>1$ ($2$- and $1$-fold degeneracy for $n=1,0$, 
respectively). Because the Hamilton of Eq. \eqref{Hamilton}
commutes with the number of excitation, it follows
that it can be diagonalized in blocks given by
\begin{align}
  &H^{(0)}=-\hbar\delta
  \nonumber\\
 &H^{(1)}=\hbar\left(
  \begin{array}{cc}
    0&g\ee^{i\theta}\sqrt{2}\\
    g\ee^{-i\theta}\sqrt{2}& -\delta
  \end{array}
 \right),
  \nonumber\\
  &H^{(n\ge 2)}=
  \nonumber
  \\
  &\hbar\left(
  \begin{array}{ccc}
    \delta+\omega(n-1)&g\ee^{i\theta}\sqrt{2(n-1)}&0\\
    g \ee^{-i\theta}\sqrt{2(n-1)}&\omega(n-1)&g\ee^{i\theta}\sqrt{2n}\\
    0&g\ee^{-i\theta}\sqrt{2n}& \omega(n-1)-\delta
  \end{array}
  \right).
  \label{Hblocks1}
\end{align}
We observe that the state $\ket{0,0}\ket0$ is an eigenstate of
the system with eigenvalue $E^{(0)}=-\hbar\delta$. For the second block we find that
there are two eigenvalues given by 
$E^{(1)}_j=\hbar\tfrac{j}{2}\left(\sqrt{8g^2+\delta^2}\right)$ with $j=-1,1$.
The solution of the eigenvalue problem  for $n\geq 2$
involves the diagonalization of the $3\times 3$
matrices $H^{(n\geq 2)}$ of Eq. \eqref{Hblocks1} and
leads to a characteristic polynomial of third order. Its general
solutions are lengthly \cite{torres} and
not of much interest for our purposes.
For large photon numbers
the approximate eigenvalues of the system can be obtained
with the help of perturbation theory.
Choosing 
$\epsilon_n=1/\sqrt{8n-4}$ as an expansion parameter for each block
we obtain the result
\begin{align}
  H^{(n)}&= H_0^{(n)}+\epsilon_n H_1^{(n)}+\dots
  \nonumber\\
  \frac{H_0^{(n)}}{\hbar}&=\left(
  \begin{array}{ccc}
    \omega(n-1)+\delta&g\ee^{i\theta}\sqrt{2n-1}&0\\
    g\ee^{-i\theta}\sqrt{2n-1}&\omega(n-1)&g\ee^{i\theta}\sqrt{2n-1}\\
    0&g\ee^{-i\theta}\sqrt{2n-1}&\omega(n-1) -\delta
  \end{array}
  \right),
  \nonumber\\
  H_1^{(n)}&=\hbar\left(
  \begin{array}{ccc}
    0&-g\ee^{i\theta}&0\\
    -g\ee^{-i\theta}&0&g\ee^{i\theta}\\
    0&g\ee^{-i\theta}&0 
  \end{array}
  \right).
  \label{Hblocksdet}
\end{align}
The eigenvalues of the zeroth order blocks are $\hbar\omega(n-1)$ and 
$\hbar\omega(n-1)\pm\hbar\Omega_n$ with 
\begin{align}
  \Omega_n=\sqrt{(4n-2)g^2+\delta^2}.
  \label{}
\end{align}
The eigenvectors of these zeroth order blocks are given by the columns of the unitary 
matrix
\begin{align}
  U_0^{(n)}
  &=\left(
  \begin{array}{ccc}
    -\frac{\omega_n\ee^{i\theta}}{\sqrt2\Omega_n}&
    \frac{\delta-\Omega_n}{2\Omega_n}\ee^{i\theta}&
    \frac{\delta+\Omega_n}{2\Omega_n}\ee^{i\theta}
    \\
    \frac{\delta}{\Omega_n}&\frac{\omega_n}{\sqrt2\Omega_n}&\frac{\omega_n}{\sqrt2\Omega_n}\\
    \frac{\omega_n\ee^{-i\theta}}{\sqrt2\Omega_n}&
    \frac{\omega_n^2\ee^{-i\theta}}{2\Omega_n(\delta-\Omega_n)}& 
    \frac{\omega_n^2\ee^{-i\theta}}{2\Omega_n(\Omega_n+\delta)}
  \end{array}
  \right).
  \label{}
\end{align}
The corrections of first order in $\epsilon_n$
of the eigenvalues are given by the diagonal
elements of the matrices $\epsilon_n{U_0^{(n)}}^\dagger H_1^{(n)} U^{(n)}_0$.
Using these corrections up to first order in $\epsilon_n$ the eigenvalues are given by
\begin{align}
  E^{(n)}_j&=\hbar
  \left(\omega(n-1)+j\Omega_n+\frac{(-1)^j(2)^{1-|j|}g^2\delta}{\Omega_n^2}\right)
  \label{Eigenvalues}
\end{align}
with $j=-1,0,1$. It should be mentioned
that these results are valid for arbitrary detunings $\delta$ from resonance. 

Let us now determine the time evolution of the two-qubit-field quantum state
with the initial condition
\begin{eqnarray}
  \ket{\Psi_0}&=&\left(
  c_-\ket{\Psi^-}+
  c_1\ket{1,1}+c_+\ket{\Psi^+}+
  c_0\ket{0,0}
  \right)\otimes \nonumber \\
&\otimes&
\left(\sum_{n=0}^{\infty}p_n\ket{n}\right)
\nonumber
\end{eqnarray} 
fulfilling the normalization condition
\begin{equation}
 \left(|c_-|^2+|c_+|^2+|c_0|^2+|c_1|^2\right)\left(\sum_{n=0}^{\infty}|p_n|^2\right)=1. \nonumber
\end{equation} 
Using the
zeroth order eigenvectors in $\epsilon_n$
and the corresponding first order eigenvalues the time  evolution 
is approximately given by 
\begin{align}
  \ket{\Psi(t)}&=
  c_-\ket{\Psi^-}\otimes
\left(\sum_{n=0}^{\infty} p_n e^{-i n\omega t}\ket{n}\right)+
  \label{psi1}\\
  &+\ket{1,1}\ket{\chi_{1}(t)}
  +\ket{\Psi^+}\ket{\chi_{0}(t)} 
  +\ket{0,0}\ket{\chi_{-1}(t)}\nonumber
\end{align}
with
 \begin{align}
  &\ket{\chi_1(t)}=
  \sum_{n=2}^\infty 
  \sum_{k=-1}^1\eta_{1,k}^{(n,t)}
  \ee^{i\theta}
  \ket{n-2},
  \nonumber\\
  &\ket{\chi_0(t)}=
  \sum_{n=1}^\infty 
  \sum_{k=-1}^1\eta_{0,k}^{(n,t)}
  \ket{n-1},
  \nonumber\\
  &\ket{\chi_{-1}(t)}=
  \sum_{n=1}^\infty 
  \sum_{k=-1}^1\eta_{-1,k}^{(n,t)}
  \ee^{-i\theta}
  \ket{n}
  +
  \ee^{i\delta t}c_0\,p_0\ket{0}
  \label{fieldstatesdelta}
\end{align}
and with the definitions
\begin{align}
  \eta_{j,\pm1}^{(n,t)}
  &=
  \tfrac{\omega_n^2}{\sqrt{2^{|j|}}\Omega_n^2}
  \left(\tfrac{\delta\pm\Omega_n}{\omega_n}\right)^j
  \Big(
  \tfrac{c_+p_{n-1}}{2}
   +
  \tfrac{
  \omega_nc_0\ee^{i\theta}p_n}
  {2\sqrt{2}(\delta\pm\Omega_n)}+
  \nonumber\\&
   \quad\quad\quad+
  \tfrac{
  (\delta\pm\Omega_n)c_1 \ee^{-i\theta} p_{n-2}}
  {2\sqrt{2}\omega_n}
  \Big)
  \ee^{-i E^{(n)}_{\pm1} t/\hbar},
  \nonumber\\
  \eta_{j,0}^{(n,t)}
  &=
  \left(\tfrac{\delta}{\omega}\right)^{\delta_{j,0}}
  (-1)^{\delta_{j,1}}
  \tfrac{\omega_n^2}{\sqrt{2^{|j|}}\Omega_n^2}
  \Big(
  \tfrac{\delta c_+p_{n-1}}{\omega_n}+ 
  \nonumber\\
  &\quad\quad\quad
   +
  \tfrac{
  c_0\ee^{i\theta}p_n-c_1\ee^{-i\theta} p_{n-2}}
  {\sqrt2}
  \Big)
  \ee^{-i E^{(n)}_0 t/\hbar}.
  \label{etast}
\end{align}
In the case of an initially prepared coherent photon state
the probability amplitudes are given by
\begin{align}
 p_n=
  \sqrt{\frac{ {\overline n}^n}{n!}}
  \ee^{-\frac{\overline n}{2}+i\phi}.
  \label{alphashort}
\end{align}
In order to obtain an expansion in terms of coherent states one may
perform a Taylor expansion of the eigenfrequencies up to first order in $n$ around
the mean photon number
$\overline n \gg 1$. 
Thus,
the eigenvalues take the form of Eq. \eqref{eigenvalues} with the definitions
of Eq. \eqref{frequenciesTaylor}. 
In the limit 
$\bar n\gg \sqrt{\bar n}$
summations over photon numbers $n$
may be restricted approximately to
intervals $n\in[\bar n-4\sqrt{\bar n},\bar n+4\sqrt{\bar n}]$. 
Thus, the probability amplitudes of the single-mode
radiation field simplify to
\begin{align*}
  p_n=\sqrt{\frac{\bar n}{n}}\ee^{i\phi}p_{n-1}\approx \ee^{i\phi}p_{n-1}
\end{align*}
and the functions of Eq. \eqref{etast} can be approximated by
\begin{align}
  \eta_{j,k}^{(n,t)}&\approx\eta_{j,k} 
  \ee^{i\left[j(\phi-(\omega+\varpi_k)(n-1)t)-\Delta_k t\right]}p_{n-j-1}. 
  \label{considereta}
\end{align}
Substituting these approximations into
Eq. \eqref{fieldstatesdelta} 
we arrive at the result of
Eq. \eqref{chistatesdelta}. Thereby, the interaction times 
are restricted by the condition of Eq. \eqref{timescale}.

\subsection{Perfect resonance $\delta=0$}
In the resonant case
the exact solutions have remarkably compact form. 
For real-valued dipole couplings, for example, i.e. $\theta=0$,
the blocks of the Hamiltonian $H^{(n)}$ 
can be diagonalized by the transformations
\begin{align}
  U^{(1)}&=\tfrac{1}{\sqrt2}\left(
 \begin{array}{cc}
   1&1\\
   -1&1
 \end{array}
  \right),
  \nonumber\\
  U^{(n\ge2)}&=\tfrac{1}{\sqrt{4n-2}}\left(
  \begin{array}{ccc}
    \sqrt{2n}&\sqrt{n-1}&\sqrt{n-1}\\
    0&-\sqrt{2n-1}&\sqrt{2n-1}\\
    \sqrt{2n-2}&\sqrt n&\sqrt n
  \end{array}
  \right).
  \label{}
\end{align}
Thereby, ${U^{(n)}}^\dagger H^{(n)}U^{(n)}$
is the diagonal matrix of eigenvalues. These eigenvalues  
are given by
$E^{n}_j=\hbar(\omega(n-1)+jg\sqrt{4n-2})$ with $j=-1,0,1$ ($j=1,-1$) for $n>1$ ($n=1$).

The resulting time evolution of an initial state 
of the form of Eq. \eqref{initial} has the form of
the state vector in equation \eqref{psi} with the field states given by
\begin{align}
  \ket{\chi_1(t)}&=\sum_{n=2}^\infty 
  \ee^{i\theta}
  \tfrac{
  \sqrt{n-1}
  \left(
  \xi_{n,t}^-
  -\xi_{n,t}^+\right)
  -\sqrt{n}\xi_{n}
  }{\sqrt{2n-1}}
  \ket{n-2},
  \nonumber\\
  \ket{\chi_0(t)}&=
  \sum_{n=1}^\infty 
  \left(
  \xi_{n,t}^-+\xi_{n,t}^+
  \right)
  \ket{n-1},
  \nonumber\\
  \ket{\chi_{-1}(t)}&=
  c_0\,p_0\ket{0}+
  \sum_{n=1}^\infty 
  \ee^{-i\theta}\tfrac{
  \sqrt{n}
  \left(
  \xi_{n,t}^-
  -\xi_{n,t}^+\right)
  +\sqrt{n-1}\xi_{n}
  }{\sqrt{2n-1}}
  \ket{n}
  \label{fieldstates}
\end{align}
with
\begin{align}
  \xi_{n,t}^\pm
  &=
  \frac{\ee^{\pm i \omega_n t}}{2}
  \left(
  c_+p_{n-1}\mp
  \tfrac{\sqrt{n}\,c_0 \ee^{i\theta}p_n+\sqrt{n-1}\,c_1 \ee^{-i\theta}p_{n-2}}{\sqrt{2n-1}}
  \right),
  \nonumber\\
  \xi_{n}
  &=
  \frac{\sqrt{n-1}\,c_0 \ee^{i\theta}p_n-\sqrt{n}\,c_1 \ee^{-i\theta}p_{n-2}}{\sqrt{2n-1}},
  \label{}
\end{align}
and
with $p_n$ denoting the photon number probability amplitudes which
are given by Eq. \eqref{alphashort}
in the case of
a coherent state.

\section{Homodyne photodetection as a projective measurement
\label{homodynedetection}}
For the sake of completeness in this appendix 
we summarize basic facts about balanced homodyne photodetection 
measurements 
which
are relevant for our discussion in Sec. \ref{Filter} and which
have been reviewed in detail by Lvovsky and Raymer \cite{Lvovsky}, for example. 
In particular, we summarize
the approximations which allow one
to describe a homodyne photodetection measurement 
by a projective von Neumann
measurement as in Eq.\eqref{homodyne}. 

In a typical balanced homodyne photodetection experiment a single mode of the radiation field to be
measured is superposed with the single mode of a local oscillator with the help of a
$50\%$ beam splitter.
Ideally this process can be described by the canonical transformation
\begin{eqnarray}
&&\left(
\begin{array}{c}
\hat{c}_2\\
\hat{c}_1
\end{array}
\right)
=
\frac{1}{\sqrt{2}}
\left(
\begin{array}{cc}
1 & 1\\
-1 & 1
\end{array}
\right) 
\left(
\begin{array}{c}
\hat{a}_S\\
\hat{a}_L
\end{array}
\right)
\end{eqnarray}
with  $\hat{a}_S$ denoting the destruction operator of the field 
mode to be measured and $\hat{a}_L$
the mode of the local oscillator.
The destruction operators of  the field modes emerging from the 
beam splitter are denoted by $\hat{c}_1$ and $\hat{c}_2$. 
With the help of two photodetectors one
measures the resulting difference of photon numbers which is described by the hermitian operator 
$\hat{n}_- = \hat{c}^{\dagger}_1\hat{c}_1 - \hat{c}^{\dagger}_2\hat{c}_2$.
According to the photodetection theory of Kelley and Kleiner 
\cite{KelleyKleiner} the probability of detecting
$n_- = n_1 - n_2$ photons 
is given by
\begin{eqnarray}
&&P(n_-) ={\rm Tr}\{
\hat{\rho}_L \otimes \hat{\rho}_S \nonumber\\
&&
:
e^{-\xi (\hat{n}_1 + \hat{n}_2)}
\left(\frac{\hat{n}_1}{\hat{n}_2}\right)^{n_-/2}
I_{|n_-|}\left( 2\xi \sqrt{\hat{n}_1 \hat{n}_2}\right)
:
\}
\label{Kelley}
\end{eqnarray}
with the mean photon numbers
\begin{displaymath}
n_j = {\rm Tr}\left\{\hat{\rho}_L
\otimes\hat{\rho_S}\, \hat{c}^{\dagger}_j \hat{c}_j\right\},~~(j=1,2).
\end{displaymath}
Thereby, it is assumed that
the two field modes described by the destruction operators 
$\hat{a}_L$
and $\hat{a}_S$ are statistically independent and are
initially prepared in the separable
quantum states $\hat{\rho}_L$ and $\hat{\rho}_S$. 
The quantity $0 \leq \xi \leq 1$ denotes the quantum 
efficiency of the photodetection process and $I_n$ denotes the
modified Bessel function of integer order $n$. Furthermore, 
normal ordering of an operator $\hat{O}$ with respect to the
destruction and creation operators $\hat{a}_j$ and 
$\hat{a}^{\dagger}_j~~(j\in\{L,S\})$ is denoted by $:\hat{O}:$.

If the magnitude of the
difference of the photon numbers $n_-$ is much less than the mean photon numbers of both modes 
emerging from the beam splitter, i.e. $| n_-| \ll n_1,n_2$, 
and in addition the local oscillator is initially prepared in a coherent state 
$\ket{|\alpha_L|e^{i\theta_L}}$ with 
$|\alpha_L|^2 \gg {\rm Tr}\left\{\hat\rho_S\hat a^\dagger_S \hat a_S \right\},1$  
this photodetection probability simplifies to the expression
\begin{eqnarray}
&&P_{\theta_L}(n_-) ={\rm Tr}
\{
\hat{\rho}_S\\
&&
:
\tfrac{1}{\sqrt{2\pi \xi |\alpha_L|^2}}
\ee^{-
\tfrac{\left(n_- - \xi |\alpha_L|(\hat{a}_Se^{-i\theta_L} + 
\hat{a}^{\dagger}_S e^{i\theta_L})
\right)^2}{2\xi |\alpha_L|^2}}
:
\}.\nonumber
\end{eqnarray}

Therefore, if the balanced homodyne detection measurement is ideal, 
i.e. $\xi \to 1$, the resulting probability
of detecting a difference photon number $n_-$ simplifies to the 
expression
\begin{equation}
P_{\theta_L}\left( q_{\theta_L}\right) =
\label{homodynegeneral1}
\int~d^2\beta~W(\beta,\beta^*) 
\delta\left(
q_{\theta_L}  - \tfrac{(\beta e^{-i\theta_L} + \beta^* e^{i\theta_L})}{\sqrt2}
\right)
\end{equation}
with  $q_{\theta_L}=n_-/\sqrt2|\alpha_L|$ and
with $W(\beta,\beta^*)$ denoting the Wigner function of the
photonic quantum state ${\hat \rho}_S$ as given by Eq. \eqref{Wignerf} and
with $\delta (x)$ denoting the Dirac delta distribution.
Using the quadrature eigenstates of Eq.\eqref{quadraturestates}
the probability distribution of Eq.\eqref{homodynegeneral1}
can be rewritten in the equivalent form of Eq.\eqref{homodyne}.
This form demonstrates explicitly that in this limit balanced
homodyne detection of photons can be described by a projective
von Neumann measurement. According to Eq.\eqref{Kelley}, however, in general
balanced homodyne detection has to be described by a positive
operator valued measure.

For a coherent state $\ket{\alpha}$, for example, the Wigner function is given by
$W(\beta,\beta^*) = 2{\rm exp}(-2\mid \beta - \alpha\mid^2)/\pi$ so that the corresponding
probability distribution of balanced homodyne photodetection is given by
\begin{equation}
P_{\theta_L}\left( q_{\theta_L}\right) = \frac{1}{\sqrt{\pi}}
{\rm exp}
\left(
-\left(q_{\theta_L} - \tilde{q}_{\theta_L}\right)^2
\right)
\end{equation}
with $\tilde{q}_{\theta_L} = 
\frac{1}{\sqrt2}(\alpha e^{-i\theta_L} + \alpha^* e^{i\theta_L})$.
Thus, postselecting photon counts by balanced homodyne photodetection
with difference photon numbers $n_-$ in the range
$n_-/\sqrt2|\alpha_L| \in (\tilde{q}_{\theta_L} - \delta_L,\tilde{q}_{\theta_L} + \delta_L)$
is equivalent to projection onto the coherent state $\ket{\alpha}$ with probability
\begin{eqnarray}
{\rm Prob}(\alpha) &=&
{\rm erf}(\delta_L) \geq  1 - \frac{e^{-\delta_L^2}}{\sqrt{\pi} \delta_L}
\label{error}
\end{eqnarray}
with ${\rm erf}(x)$ denoting the error function \cite{AS}. Thus, choosing $\delta_L = 2$,
for example, yields ${\rm Prob}(\alpha) >  0.9953222650$ and $\delta_L = 3$ yields
${\rm Prob}(\alpha) > 0.9999779095$.

\section{Time evolution of different qubits}
\label{DiffAt}
In this appendix we analyze the more general
situation  when the qubits have different
coupling strengths to the cavity and also different transition frequencies.
We focus on small deviations from the ideal Hamiltonian in Eq. \eqref{Hamilton} which in 
this case is replaced by
\begin{align}
  \hspace{-.23cm}\hat{H}&= 
\hbar \omega \hat{a}^\dagger \hat{a} 
+\hspace{-.21cm}\sum_{i=A,B}\hspace{-.13cm}\hbar\left( 
\frac{\delta_i-\omega}{2}\hat\sigma_i^z
+ g_i  \hat{\sigma}^+_i\hat{a}+ g_i 
  \hat{\sigma}^-_i\hat{a}^\dagger\right)
\label{Hamiltoneps}
\end{align}
where $g_A=g+\varepsilon_g$, $g_B=g-\varepsilon_g$, 
$\delta_A=\delta+\varepsilon_\delta$
and $\delta_B=\delta-\varepsilon_\delta$. 
The state $\ket{\Psi^-}\ket n$ is no longer an eigenstate of the Hamiltonian
in Eq. \eqref{Hamiltoneps}.
Therefore, for photon numbers $n\geq 2$ its blocks are  
$4\times 4$ matrices which  in the basis 
$\{\ket{\psi^-}\ket{n-1},\ket{1,1}\ket{n-2},\ket{\Psi^+}\ket{n-1},
\ket{0,0}\ket{n}\}$  
are given by
\begin{align}
  \label{Hblockeps}
  &H^{(n\ge 2)}=\\
  &\hbar\left(
  \begin{array}{cccc}
    \omega(n-1) & \varepsilon_g \sqrt{2n-2}& -\varepsilon_\delta & -\varepsilon_g\sqrt{2n}\\
    \varepsilon_g\sqrt{2n-2} &\omega(n-1)+\delta& g\sqrt{2n-2} & 0\\
    -\varepsilon_\delta & g\sqrt{2n-2} & \omega(n-1) & -g\sqrt{2n}\\
     \varepsilon_g\sqrt{2n} & 0 & -g\sqrt{2n} & \omega(n-1)-\delta
  \end{array}
  \right).\nonumber
\end{align}
Thereby, for the sake of simplicity we have concentrated to the special case
of $\theta = 0$ so that the coupling strengths $g$ are positive. 
However, the eigenvalues of the matrices of Eq. \eqref{Hblockeps}
do not depend on the choice of this phase.

If $\delta\ll \omega_{\overline n}$ and the number 
of excitations is large, i.e. $n>>1$,
the four eigenvalues of each block are 
approximately  given by the two pairs  
$\hbar\omega(n-1)\pm\hbar \Omega^{(\rm S)}_n$ 
and $\hbar\omega(n-1)\pm\hbar \Omega^{(\rm L)}_n$. They reduce to
$\hbar \omega(n-1)\pm 0$ and $\hbar \omega (n-1)\pm \hbar\Omega_n$
in the limit of $\epsilon_g\to0$ and $\epsilon_\delta\to 0$.
Therefore, the asymmetries between the coupling strengths  and the detunings
induce new Rabi oscillations that will collapse (and revive)  at a slower
time scale. This behaviour can be  identified in Fig. \ref{overlapeps}.
To estimate this time scale we aim for a coherent state expansion
of the photonic state $\ket{\chi_2(t)}$ involving the smallest frequencies. 
In order to obtain simple analytical solutions that approximate the eigenvalues
of \eqref{Hblockeps} we take the zeroth order expansion analogue to 
Eq. \eqref{Hblocksdet}.
Linearizing the pair of smallest eigenvalues in $\varepsilon_g$, 
$\varepsilon_\delta$ and $n-\overline n$ we obtain
$\Omega_n^{(\rm S)}\approx\Delta+\varpi (n-\overline n)$ with
\begin{align}
  \Delta&=\frac{\omega_{\overline n}^2}{g\Omega_{\overline n}}\varepsilon_g
  +\frac{\delta}{\Omega_{\overline n}}\varepsilon_\delta,
  \nonumber\\
  \varpi&=
  2\frac{\Omega_{\overline n}^2+\delta^2}{\Omega_{\overline n}^3}g \varepsilon_g
  +
  \frac{g^2\delta}{\Omega_{\overline n}^3}\varepsilon_\delta.
  \label{freqeps}
\end{align}

The field state that accompanies the Bell state $\ket{\Psi^-}$ in Eq. \eqref{psieps} 
can be assumed to depend on coherent states in 
a similar way as the states in Eq. \eqref{chistatesdelta} and therefore can
be written in the form
\begin{align}
  \ket{\chi_2(t)}\approx \sum_\pm \eta_{2,\pm}\ee^{\mp i(\Delta-\varpi \overline n) t}
  \ket{\alpha \ee^{-i(\omega\pm \varpi)t}}.
  \label{chi2eps}
\end{align}
In the case of small values of $\varepsilon_\delta$ and $\varepsilon_g$ 
the deviation of the state $\ket{\chi_2(t)}$ 
from the coherent state evolving with frequency $\omega$ is small.
The exact form of the coefficients in Eq. \eqref{chi2eps} is not relevant
for our analysis as we focus only on the frequencies of the system. 
From Eq. \eqref{overlap} we can conclude that the overlap 
$|\braket{\alpha \ee^{-i\omega t}}{\chi_2(t)}|^2$ undergoes Rabi oscillations
at frequency  $2\Delta$ which decay as $\exp\{-\overline n\varpi^2 t^2\}$.

At this point it is convenient to summarize the steps of our scheme in order
to have a clear picture of how the fidelity of the final state and
the success probability of post-selecting the state $\ket{\Psi^-}$ change. 
We start with the initial state of Eq. \eqref{initial} which
for an interaction time $\tau$
evolves
under the influence of
the Hamiltonian in Eq. \eqref{Hamiltoneps} to a state 
given by Eq. \eqref{psieps}. A projection onto the photonic state 
$\ket{\alpha \ee^{-i\omega\tau}}$ is performed with success probability
$P=\sum_{j=-1}^2|z_j|^2$ which is given in terms of the overlaps $z_j=
  \braket{\alpha\ee^{-i\omega\tau}}{\chi_j(\tau)}$.
In the ideal case this reduces to the expression of Eq. \eqref{Prob}. The resulting
material qubits are allowed to interact with a second cavity prepared
in a coherent state that differs from the first coherent state by
a phase $\tilde \varphi$  given in Eqs. \eqref{tildevarphi} and
\eqref{condition}.
Thereby, the tripartite system for the second interaction is given by
\begin{displaymath}
  \ket{\Psi_1}=
  \tfrac{1}{\sqrt{P}}
  \left(
  z_2\ket{\Psi^-}+z_{1}\ket{1,1}+z_0 \ket{\Psi^+}
  +z_{-1}\ket{0,0}
  \right)\ket{\alpha \ee^{i\tilde \varphi}}.
\end{displaymath}
This state evolves under the action of a Hamiltonian $\hat H'$ in the form
of Eq. \eqref{Hamiltoneps}
to the state $\ket{\Psi'(\tau)}$ in the form
of Eq. \eqref{psieps} with the photonic states $\ket{\chi_j'(\tau)}$, $j=-1,0,1,2$. 
A projection  onto the field state $\ket{\alpha \ee^{i(\tilde\varphi-\omega\tau)}}$
is performed with success probability
$P'=\sum_{j=-1}^2
  |
  z_j'
  |^2$ with
  $z_j'=
  \braket{\alpha\ee^{i(\tilde \varphi-\omega\tau)}}{\chi'_j(\tau)}$. 
The total success probability of the scheme is given by $P_T=P P'$ 
and the material qubits result in the state
  \begin{displaymath}
    \ket{\psi_f}=\tfrac{1}{\sqrt{ P'}}\left(
  z_2'\ket{\Psi^-}+z_{1}'\ket{1,1}+z_0' \ket{\Psi^+}
  +z_{-1}'\ket{0,0}
  \right)
    \label{}
  \end{displaymath}  
which in the ideal case matches the Bell state $\ket{\Psi^-}$. 
The  fidelity with respect to this Bell state 
is $F=|\braket{\Psi^-}{\psi_f}|$ and in the ideal case
it attains values close to unity.

In our numerical examples
the success probability involves two measurements at time $\tau\to\tau_r/4$
with the revival time $\tau_r$ as given by Eq. \eqref{revcoltime}.
Its decay depends on $\varpi$ and it
is proportional to $\exp\{-2\overline n\varpi^2 (\tau_r/4)^2\}$. 
This sets a boundary for the applicability of our scheme as the success probabilities
become arbitrarily small for large arguments of the exponential function. 
The boundary for which
the argument of the exponential is
less than unity is determined by the inequality
\begin{align}
  \frac{\varepsilon_g}{g}
  \le
  \frac{1}{\pi}
  \sqrt{\frac{2}{\overline n}}
  \left(1+\frac{\delta^2}{\Omega_{\overline n}^2}\right)^{-1}
  -
  \frac{1}{2}
  \frac{\delta\varepsilon_\delta}{\Omega_{\overline n}^2+\delta^2}.
  \label{boundeps}
\end{align}


The oscillations of the fidelity in Fig. \ref{fidsuc} achieve their maximum
values close to unity whenever the asymmetry $\varepsilon_g$ is such that the
corresponding  Rabi oscillation completes a cycle at the interaction
time $\tau=\tau_r/4$.
This condition is fulfilled for
$2\Delta\tau_r/4=2 l \pi$ with $l\in \mathbb{N}_+$. 
From Eq. \eqref{freqeps} we obtain that this happens for 
\begin{align}
\varepsilon_g^{(l)}/g=
  \left(4g^2l-\delta\varepsilon_\delta\right)
  /\omega_{\overline n}^2
  \label{}
\end{align}
which is taken for a fixed value of $\varepsilon_\delta$. 
The first minimum in the fidelity as a function of $\varepsilon_g$ occurs at 
$\varepsilon_g^{(1)}/2$. For $\varepsilon_\delta=0$ this
value is approximately $1/2\overline n$.



\begin{thebibliography}{99}
\bibitem{footnote}
We assume that the state of the single mode resonator can be perfectly 
transferred into a single signal mode to be measured by  balanced homodyne
photodetection.
\end{thebibliography}
\end{document}